\documentclass[%
superscriptaddress,
preprint,
 amsmath,amssymb,
 aps,
]{revtex4-1}

\usepackage{graphicx}
\usepackage{dcolumn}
\usepackage{bm}

\usepackage{notes2bib}
\bibnotesetup{
note-name = ,
use-sort-key = false,
placement = mixed
}

\usepackage{xcolor}

\newcommand{\jgr}{    {J. Geophys. Res.}}
\newcommand{\ssr}{    {Space Sci. Rev.}}

\newcommand{\blue}{\textcolor{black}}

\begin{document}

\preprint{}

\title{Cross-scale energy transfer from fluid-scale Alfv\'en waves to kinetic-scale ion acoustic waves in the Earth's magnetopause boundary layer}

\author{Xin An}
\email[]{phyax@ucla.edu}
\homepage[]{https://sites.google.com/view/xin-an-physics}
\affiliation{Department of Earth, Planetary, and Space Sciences, University of California, Los Angeles, CA, 90095, USA}

\author{Anton Artemyev}
\affiliation{Department of Earth, Planetary, and Space Sciences, University of California, Los Angeles, CA, 90095, USA}

\author{Vassilis Angelopoulos}
\affiliation{Department of Earth, Planetary, and Space Sciences, University of California, Los Angeles, CA, 90095, USA}

\author{Terry Z. Liu}
\affiliation{Department of Earth, Planetary, and Space Sciences, University of California, Los Angeles, CA, 90095, USA}

\author{Ivan Vasko}
\affiliation{Department of Physics, University of Texas at Dallas, Richardson, TX, 75080, USA}

\author{David Malaspina}
\affiliation{Astrophysical and Planetary Sciences Department, University of Colorado, Boulder, CO, 80305, USA}
\affiliation{Laboratory for Atmospheric and Space Physics, University of Colorado, Boulder, CO, 80303, USA}

\date{\today}

\begin{abstract}
In space plasmas, large-amplitude Alfv\'en waves can drive compressive perturbations, accelerate ion beams, and lead to plasma heating and the excitation of ion acoustic waves at kinetic scales. This energy channelling from fluid to kinetic scales represents a complementary path to the classical turbulent cascade. Here, we present observational and computational evidence to validate this hypothesis by simultaneously resolving the fluid-scale Alfv\'en waves, kinetic-scale ion acoustic waves, and their imprints on ion velocity distributions in the Earth's magnetopause boundary layer. We show that two coexisting compressive modes, driven by the magnetic pressure gradients of Alfv\'en waves, not only accelerate the ion tail population to the Alfv\'en velocity, but also heat the ion core population near the ion acoustic velocity and generate Debye-scale ion acoustic waves. Thus, Alfv\'en-acoustic energy channeling emerges as a viable mechanism for plasma heating near plasma boundaries where large-amplitude Alfv\'en waves are present.
\end{abstract}

\maketitle


Understanding the multiscale energy transfer from magnetohydrodynamic electromagnetic perturbations and convective flows to kinetic field structures and plasma heating is of crucial importance in space and many other plasmas. The classical picture suggests that turbulent energy cascades from large to successively smaller scales through nonlinear interactions between counter-propagating Alfv\'en waves \cite{Iroshnikov64,Kraichnan65,Goldreich&Sridhar97,boldyrev2006spectrum,howes2012toward}. A complementary path of energy transfer involves the electrostatic coupling between Alfv\'enic and ion acoustic fluctuations. In this scenario, magnetic pressure gradients of Alfv\'en waves induce density perturbations and electric fields parallel to the background magnetic field \cite[e.g.,][]{hollweg1971density}, which accelerate ion beams through nonlinear Landau resonance \cite{medvedev1998asymptotic,araneda2008proton,matteini2010kinetics}. These ion beams then excite kinetic-scale ion acoustic waves (IAWs), which in turn relax ion beams, leading to parallel ion heating and terminating the energy transfer \cite{valentini2008cross,valentini2009electrostatic,valentini2011short,valentini2014nonlinear}. The implications of Alfv\'en-acoustic channeling on multiscale energy transfer span across space, astrophysical, and fusion plasmas. In the solar wind, IAWs and associated ion beams are observed around magnetic discontinuities or switchbacks \cite{graham2021kinetic,mozer2020large,malaspina2023statistical,malaspina2024frequency}, which likely evolve from outward-propagating Alfv\'en waves \cite{squire2020situ,mallet2021evolution,tenerani2021evolution}, resulting in significant ion heating \cite{mozer2022core,kellogg2024heating,woodham2021enhanced}. In Tokamak plasmas, Alfv\'en waves driven by suprathermal ions transfer some of their energy to IAWs, which are heavily Landau-damped by thermal ions, thereby heating the thermal ions \cite{gorelenkov2009beta,curran2012low,bierwage2015alfven,chen2016physics}.

Despite the importance of Alfv\'en-acoustic energy channeling, direct observational evidence of this process has been difficult to obtain, mainly because of the instrumentation required to resolve ion velocity distributions at a high time cadence, as well as the vast scale separation between Alfv\'en waves and IAWs.
In this Letter, we test the hypothesis of electrostatic-driven, cross-scale energy transfer from Alfv\'en waves to IAWs using the Earth's magnetopause boundary layer as a natural laboratory, where the solar wind interaction with magnetosphere generates a plethora of large-amplitude Alfv\'en waves. The four-satellite Magnetospheric Multiscale (MMS) mission \cite{Burch16} provides simultaneous measurements for fluid-scale Alfv\'en waves (spatial scales resolved by inter-spacecraft interferometry), kinetic-scale IAWs (spatial scales resolved by inter-antenna interferometry with a single spacecraft), and ion velocity distributions (enabled by 3D measurements of ion velocity space with a high time resolution). The interpretation of these data is supported by event-oriented kinetic simulations.

On 8 September 2015, the MMS constellation traversed from the duskside magnetosphere to the magnetosheath between 09:10 and 11:40\,UT. A surface wave, generated by the Kelvin-Helmholtz instability, is seen through repetitive crossings of current sheets, which separate the relatively cold, dense magnetosheath from the hot, tenuous magnetopause boundary layer \cite{eriksson2016magnetospheric}. Thanks to the long interval ($\sim 80$ minutes) of continuous burst-mode data collected during boundary layer crossing, a rich variety of complex plasma dynamics was observed, including Alfv\'enic turbulence \cite{stawarz2016observations}, ion beams and plasma heating \cite{sorriso2019turbulence}, and large-amplitude ($\sim 100$\,mV/m) electrostatic waves \cite{wilder2016observations}. 

Figure \ref{fig:MMS} shows an example of magnetopause crossings between 10:35:35 and 10:36:05\,UT \blue{(see Supplemental Materials \bibnote{See Supplemental Materials for the repetitive crossings of the magnetopause boundary layer over the $2.5$ hours interval, which include Refs.~\cite{Burch16,Russell16:mms,Ergun16:ssr,Lindqvist16,Pollock16:mms}.} for the whole event of magnetopause crossing)}. Within the magnetopause boundary layer, enhanced Alfv\'en waves with a normalized amplitude $\vert \delta B / B_0 \vert \sim 0.2$ are identified by the correlated perturbations between magnetic and velocity fields, as shown in Figure \ref{fig:MMS}(f). Here $B_0$ is the background magnetic field averaged in the magnetopause boundary layer. The magnetic field data from the four MMS spacecraft show clear time shifts among them, due to wave propagation and plasma flows. Using four-spacecraft interferometry \blue{(see explanations in Supplemental Materials \bibnote{See the four-spacecraft interferometry analysis of Alfv\'en waves in Supplemental Materials, which include Ref.~\cite{paschmann2000issi}})}, we determine the parallel wave propagation velocity to be $539\,\mathrm{km/s} \approx v_\mathrm{A}$ in the plasma rest frame and the corresponding wavelength to be $2059\,\mathrm{km} \approx 30\, d_i$ for the dominant frequency $0.05\,\mathrm{Hz} = 0.05 f_{ci}$ in the spacecraft frame ($0.26\,\mathrm{Hz} = 0.26 f_{ci}$ in the plasma rest frame). Here, $d_i = 68$\,km, $f_{ci} = 1$\,Hz and $v_\mathrm{A} = 506$\,km/s are the ion inertial length, ion gyrofrequency and Alfv\'en velocity, respectively, averaged over the interval 10:35:40--10:36:00\,UT. These large-amplitude Alfv\'en waves steepen into current sheets with strong gradients in the total magnetic field (e.g., around 10:35:46\,UT), which are colocated with density bumps and low-frequency parallel electric fields [Figures \ref{fig:MMS}(b) and \ref{fig:MMS}(c)]. Such density and electric field perturbations are likely driven by gradients in the wave magnetic-field pressure [Figure \ref{fig:MMS}(c)], and can be viewed as the ion acoustic mode in the long-wavelength limit \cite{hollweg1971density}.

\begin{figure}[h!]
    \centering
    \includegraphics[width=\textwidth]{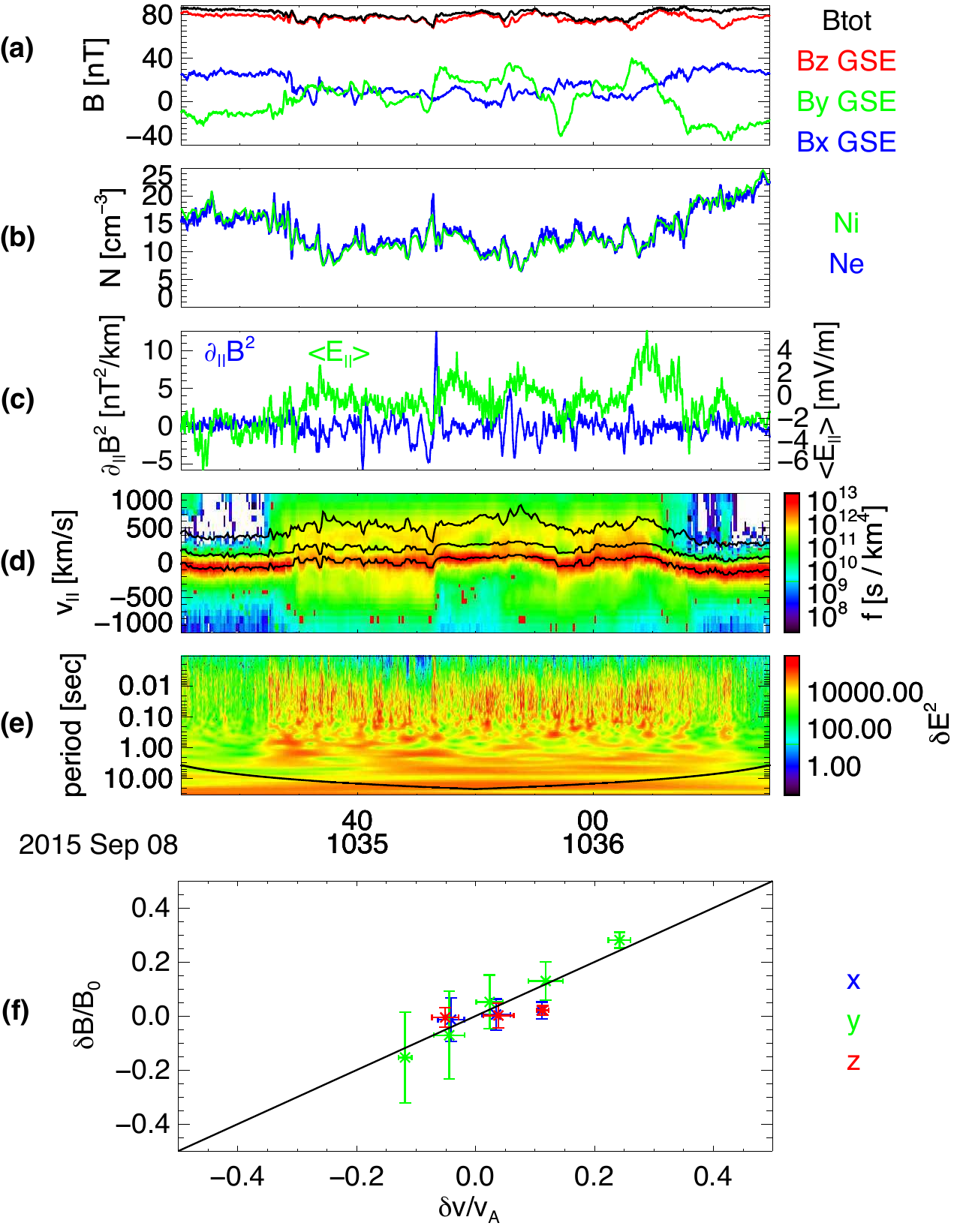}
\end{figure}
\begin{figure}[h!]
    \caption{An example of magnetopause boundary layer crossing by MMS1 on 8 September 2015. The boundary layer is observed between 10:35:33 and 10:36:08\,UT. (a) Three components of magnetic field in Geocentric Solar Ecliptic (GSE) coordinate system measured by the Fluxgate Magnetometer at $128$\,Samples/second \cite{Russell16:mms}. The total magnetic field strength is shown in black. (b) Ion and electron densities measured by the Fast Plasma Investigation (FPI) instrument at time cadences of $150$\,ms and $30$\,ms, respectively \cite{Pollock16:mms}. (c) Parallel spatial gradients of magnetic field pressure and smoothed parallel electric fields. (d) Reduced ion parallel velocity distributions obtained from integrating phase space densities measured in the 3D velocity space (energy, pitch angle, gyrophase) by the FPI instrument. The phase space density is coded in color. The three solid lines from top to bottom indicate Alfv\'en and ion acoustic velocities relative to the ion core, and the maximum phase space density at each time, respectively. (e) Wavelet analysis of electric field measured by the Electric Double Probes at a sampling rate $8192$\,Samples/second \cite{Lindqvist16,Ergun16:ssr}. The spectral density is coded in color. The black line tracks the cone of influence, below which the stretched wavelets extend beyond the edges of the observation interval. (f) Correlated magnetic and plasma flow perturbations for the interval 10:35:35--10:36:05\,UT. The magnetic field is normalized by the mean magnetic field of this interval. The plasma flow velocity is normalized by the mean Alfv\'en velocity of this interval. The $x$, $y$ and $z$ components of each perturbation is color-coded in blue, green, and red.}
    \label{fig:MMS}
\end{figure}

Ion beams (local maximums in phase space density in $v_\parallel$ aside from the ion core) are seen in the reduced velocity distributions at variable parallel velocities up to $v_\mathrm{A} = 506$\,km/s relative to the ion core [Figure \ref{fig:MMS}(d)]. These ion beams are believed to be driven by resonant interactions between ions and Alfv\'enic fluctuations \cite{sorriso2019turbulence}. Intense broadband electrostatic waves appear during each boundary layer crossing throughout the $\sim 2$-hour event [Figure \ref{fig:MMS}(e)], with frequencies ranging from $10$\,Hz to the ion plasma frequency $f_{pi} = 700$\,Hz. Identified as the ion acoustic mode, these waves are likely excited by the ion beams \cite{wilder2016observations}. By analyzing time delays between the voltage signals in each pair of opposing voltage-sensitive probes in three orthogonal directions (with tip-to-tip effective distances of $120$\,m in the spin plan and $30$\,m along the spin axis), we perform interferometry for short-wavelength IAWs at various locations in the boundary layer \blue{(see analysis in Supplemental Materials \bibnote{see Supplemental Materials for the detailed inter-antenna interferometry analysis, which include Refs.~\cite{vasko2018solitary,vasko2020nature,Sonnerup68}})}. The IAWs propagate within $15^\circ$ relative to $B_0$ or $-B_0$, with phase speeds $0.1$--$1.5\, c_s$ in the plasma rest frame. They have wavelengths of $1$--$30\, \lambda_D$, where $c_s$ is the ion acoustic velocity and $\lambda_D$ the local Debye length. The antiparallel-propagating IAWs are consistent with ion beams in the antiparallel direction, although these beams are weaker than those in the parallel direction.

It is important to determine the ordering of ion acoustic and Alfv\'en velocities, $c_s$ and $v_\mathrm{A}$, relative to the thermal velocity of core ions $v_\mathrm{Ti}$ (excluding beam ions). The electron-to-ion temperature ratio is $T_e / T_i \approx 0.25$, and the ion beta is $\beta_i = n_0 T_i / (B_0^2 / 8 \pi) \approx 0.15$, which yields $c_s / v_\mathrm{Ti} = \sqrt{(5/3) + (T_e / T_i)} \approx 1.4$ and $v_\mathrm{A} / v_\mathrm{Ti} = \sqrt{2/\beta_i} \approx 3.7$. Here, $n_0$ is the background density averaged in the magnetopause boundary layer. Thus, the ion beams at $v_\mathrm{A} = 3.7 v_\mathrm{Ti} = 506$\,km/s (i.e., the tail of the velocity distribution) cannot be in Landau-resonance with the IAWs propagating along the field at $c_s = 1.4 v_\mathrm{Ti} = 191$\,km/s (i.e., the core of the velocity distribution). Nevertheless, there must be some ion beams present at $c_s$ to drive the IAWs; otherwise, those waves would be heavily damped by thermal core ions through Landau damping. This expectation is strongly supported by the presence of ion beams at $\sim 200$\,km/s relative to the ion core in Figure \ref{fig:MMS}(d).

To facilitate the interpretation of \textit{in situ} observation data, we perform event-oriented kinetic simulations using the Hybrid-VPIC code \cite{bowers2008ultrahigh,le2023hybrid}. This code treats ions as kinetic particles and electrons as a massless fluid. Our simulation spans two dimensions $[0 \leq x \leq 120\,d_i, 0 \leq y \leq 15\,d_i]$ in configuration space and three dimensions $(v_x, v_y, v_z)$ in velocity space. The cell size is $\Delta x = \Delta y = 0.059 d_i$. The time step is $\Delta t = 0.004 \omega_{ci}^{-1}$ for particle push, which is divided to $10$ substeps for the field solver to satisfy the Courant condition $(\Delta t / 10) \cdot \omega_{ci} < (\Delta x / d_i)^2 / \pi$ for the fastest eigenmode, the whistler mode. Periodic boundary conditions are applied to both fields and particles. All numerical values for fields and particles are based on MMS observations in Figure \ref{fig:MMS}. A uniform background magnetic field $B_0$ is applied along the $+x$ direction. Initially, the system is perturbed by a spectrum of parallel-propagating Alfv\'en waves prescribed by $\delta B_x = 0$, $\delta B_y = -\delta B_{\perp} \sum_{m = 3}^{5} \sin (2 \pi m x / L_x + \phi_i)$, and $\delta B_z = \delta B_{\perp} \sum_{m = 3}^{5} \cos (2 \pi m x / L_x + \phi_i)$, where $\delta B_\perp / B_0 = 0.15$ is the wave amplitude, $L_x$ is the system size in the $x$ direction, and $m$ is the mode number. The initial wave phases are uniformly distributed between $0$ and $2\pi$ as $\phi_1 = 0$, $\phi_2 = 2 \pi /3$, and $\phi_3 = 4 \pi / 3$. Ions are initialized as a drifting Maxwellian with thermal velocity $v_\mathrm{Ti} = 0.27\, v_\mathrm{A}$ and perturbed transverse fluid velocities commensurate with Alfv\'en waves, $\delta v_x = 0$, $\delta v_y = \delta v_{\perp} \sum_{m = 3}^{5} \sin (2 \pi m x / L_x + \phi_i)$, and $\delta v_z = - \delta v_{\perp} \sum_{m = 3}^{5} \cos (2 \pi m x / L_x + \phi_i)$, where $\delta v_{\perp} / v_\mathrm{A} \approx \delta B_\perp / B_0 = 0.15$. The uniform ion density $n_0$ at $t=0$ is sampled by $400$ particles in each cell. The electron-to-ion temperature ratio is $0.25$. Because the electron thermal velocity is much greater than the Alfv\'en and ion acoustic velocities, the equation of state for electrons is isothermal, i.e., $T_e = \mathrm{constant}$. The results are presented in normalized units: time to $\omega_{ci}^{-1}$, lengths to $d_i$, velocities to $v_\mathrm{A}$, densities to $n_0$, electric fields to $v_\mathrm{A} B_0 / c$ (where $c$ is the speed of light), and magnetic fields to $B_0$.

\begin{figure}[h!]
    \centering
    \includegraphics[width=\textwidth]{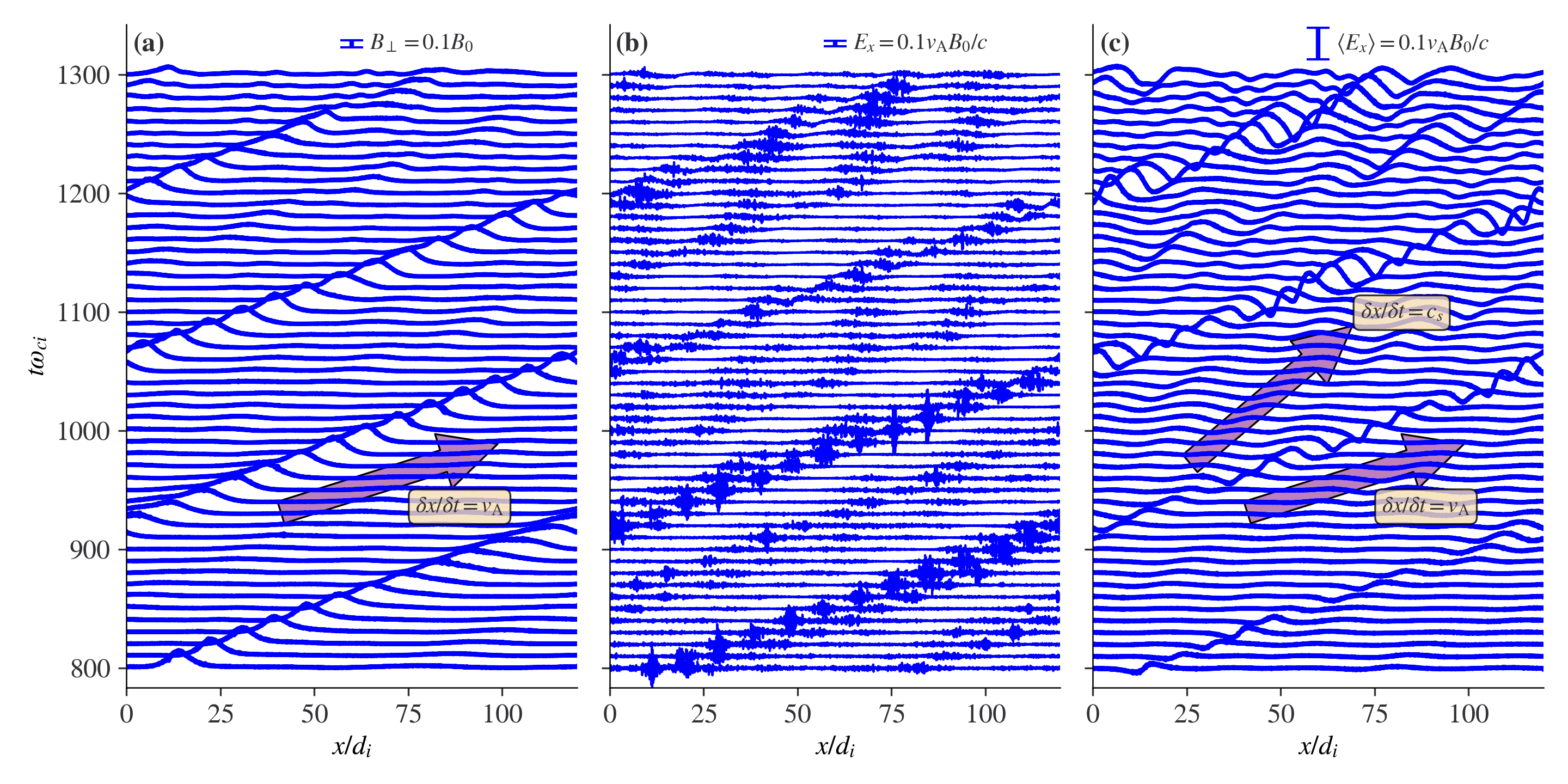}
    \caption{Propagation characteristics of Alfv\'en waves and IAWs in the spatiotemporal domain in the simulation. The horizontal axis represents the $x$ direction parallel to $B_0$. The vertical axis represents time. The amplitudes of oscillations are shown at the top of each panel. (a) The perpendicular magnetic field $\delta B_\perp$. (b) The full parallel electric field $\delta E_x$. (c) The parallel electric field $\langle \delta E_x \rangle$ averaged over the $y$ direction. The short-wavelength IAWs are canceled out by averaging in the $y$ direction, whereas the long-wavelength electric fields survive.}
    \label{fig:hybrid-propagation}
\end{figure}

\begin{figure}[h!]
    \centering
    \includegraphics[width=\textwidth]{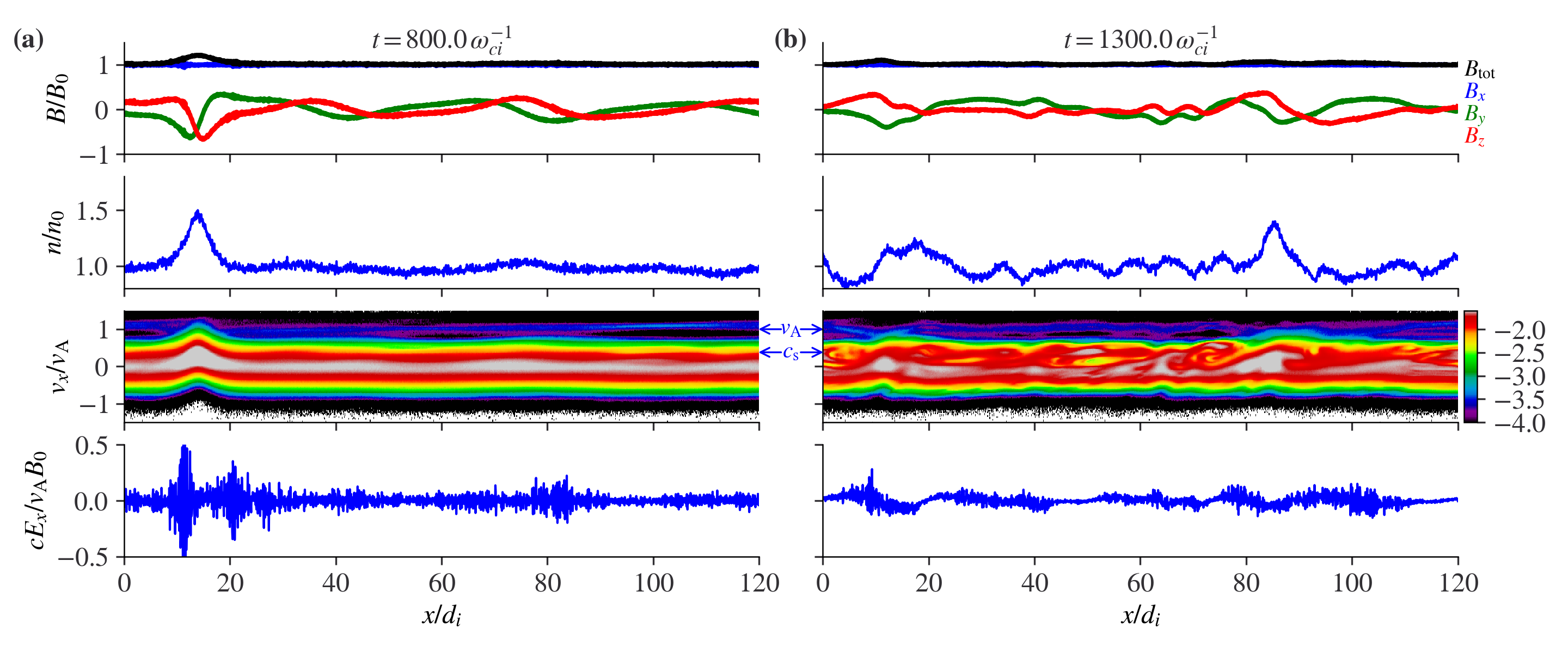}
    \caption{The steepened Alfv\'en waves and the associated ion beams and kinetic-scale IAWs in the simulation. The shown time snapshots are at (a) $t = 800\,\omega_{ci}^{-1}$ and (b) $t = 1300\,\omega_{ci}^{-1}$. In each snapshot, the four panels from top to bottom are (1) the three components of magnetic field and the total strength, (2) the plasma density, (3) the ion phase portrait color-coded by the phase space density, and (4) the parallel electric field, respectively. The horizontal axis for all panels are the $x$ direction parallel to $B_0$.}
    \label{fig:hybrid-phasespace}
\end{figure}

The large-amplitude, broadband Alfv\'en waves naturally exhibit magnitude modulations, leading to phase-steepened wavefronts (see Figure \ref{fig:hybrid-propagation}(a) and Refs.~\cite{cohen1974nonlinear,gonzalez2021proton}). At these steepened wavefronts, rotational discontinuities are observed in the rapid phase changes of $B_y$ and $B_z$ over a distance of $\sim 5 d_i$ [Figure \ref{fig:hybrid-phasespace}(a)]. The gradients of magnetic pressure drive compressive perturbations, as described by \cite{hollweg1971density}
\begin{equation}\label{eq:density-perturb}
    \frac{\partial^2 \delta \rho}{\partial t^2} - c_s^2 \frac{\partial^2 \delta \rho}{\partial x^2} = \frac{1}{8 \pi} \frac{\partial (\delta B_\perp)^2}{\partial x} ,
\end{equation}
where $\delta \rho$ is the mass density perturbation, and $\delta B_\perp$ is the perpendicular magnetic field of the Alfv\'en waves. Consequently, a density bump of $\delta \rho / \rho_0 \lesssim 0.5$ forms at the steepened wavefront, accompanied by a parallel ion flow with a streaming velocity $\delta v / v_\mathrm{A} = \delta \rho / \rho_0 \lesssim 0.5$ [at $x \approx 15\,d_i$ in Figures \ref{fig:hybrid-phasespace}(a)]. The relative streaming between core ions and electrons along the magnetic field are unstable to the Buneman or current-driven instability \cite{buneman1959dissipation,davidson1970electron}, and are expected to generate IAWs with a maximum growth rate at the wavelength $\lambda \sim c_s / \omega_{pi} = \lambda_D \sqrt{(5T_i/3T_e) + 1} = 2.7 \lambda_D$. Indeed, short-wavelength IAWs are observed near the steepened wavefront, appearing as short packets with spiky electric fields [Figure \ref{fig:hybrid-phasespace}(a)]. The expected wavelength is consistent with that measured by MMS ($1$--$30\,\lambda_D$) \bibnote{IAWs are allowed to propagate in the hybrid-kinetic scheme. However, the Debye-scale IAWs cannot be resolved in this scheme, because there is no other intrinsic, physical spatial scale in the system smaller than the ion inertial length.}.


Equation~\eqref{eq:density-perturb} reveals two compressive modes driven by large-amplitude Alfv\'en waves. The first mode is a particular solution to the equation, which is localized and attached to the steepened wavefront propagating at $v_\mathrm{A}$. The second mode is a solution to the homogeneous part of the equation, which is periodically emitted by the steepend wavefront once a density perturbation is established, propagating at $c_s$. These two modes are observed in the long-wavelength electric fields [Figure \ref{fig:hybrid-propagation}(c)].

The two compressive modes are also manifested in the ion phase portrait. First, the localized electrostatic field at the steepened wavefront accelerates the ion tail population to $v_\mathrm{A}$ through Landau resonance (see Figure \ref{fig:hybrid-phasespace}(a) and Refs.~\cite{araneda2008proton,valentini2008cross,matteini2010kinetics,gonzalez2023particle}). Second, the periodic electrostatic field propagating at $c_s = 1.4\, v_\mathrm{Ti} = 0.38\, v_\mathrm{A}$ traps the ion thermal population, forming beams in resonant islands through nonlinear Landau resonance (see Figure \ref{fig:hybrid-phasespace}(b) and Ref.~\cite{gonzalez2023particle}). These beams then excite short-wavelength IAWs via the ion-ion two-stream instability, with phase velocities ranging from $c_s - \Delta v_{\mathrm{trap}}$ to $c_s + \Delta v_{\mathrm{trap}}$, where $\Delta v_{\mathrm{trap}}$ is the resonant island half-width. These propagation velocities align with MMS interferometry analysis. Comparing Figures \ref{fig:hybrid-phasespace}(a) and \ref{fig:hybrid-phasespace}(b) shows that the IAW amplitude from the current-driven (Buneman) instability near the steepened wavefront significantly exceeds that of IAWs excited by the ion-ion two-stream instability elsewhere. Asymptotically, ion beams within resonant islands phase-mix to a plateau distribution, indicating resonant heating by Alfv\'en waves' nonlinearly generated parallel electric fields. The short-wavelength IAWs persist due to this plateau distribution \cite{valentini2008cross,valentini2009electrostatic}, analogous to electron-acoustic solitary structures in the magnetosphere \cite{holloway1991undamped,valentini2006excitation,An19,An21:kaw}. 

A comparable process linking large-scale electromagnetic and small-scale electrostatic phenomena occurs in the auroral ionosphere \cite{Genot01,Genot04}. In this context, Alfv\'en wave propagation through inhomogeneous auroral plasma generates parallel electric fields, potentially contributing to similar field generation in our case. Additionally, hydrodynamic filamentation instability \cite{mottez1992filamentation,gedalin2010growth}, arising from relative streaming between beams and background plasma, can lead to the development of smaller spatial scales within the fluid regime. However, the ultimate excitation of Debye-scale electric fields necessitates kinetic instabilities, as demonstrated in our analysis.

\begin{figure}[h!]
    \centering
    \includegraphics[width=\textwidth]{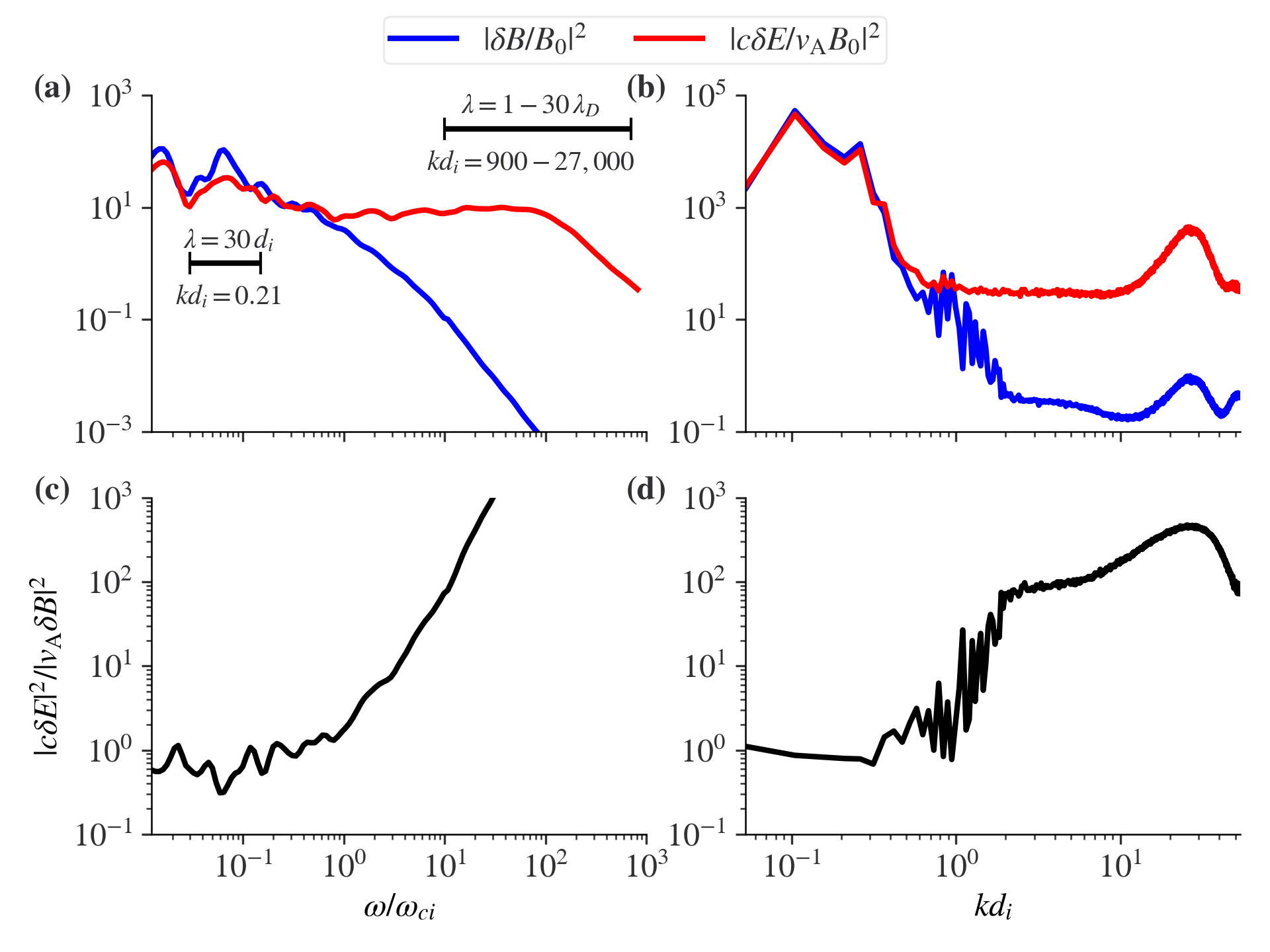}
    \caption{Comparison of electromagnetic power spectra between the MMS observation and the simulation. (a) The electric and magnetic power spectra as functions of wave frequency and (c) the ratio between the two spectra from the MMS observation in Figure \ref{fig:MMS}. Note that the shown wave frequency is in the spacecraft frame, which is Doppler-shifted from that in the plasma rest frame. There is no simple conversion between the two frequencies, because the amount of Doppler shift varies with wave frequency or wavenumber. To facilitate the mapping between wave frequencies in the spacecraft frame and wavenumbers in the plasma frame, we mark the wavenumbers of Alfv\'en waves and IAWs at their corresponding frequencies in the spacecraft frame. (b) The electric and magnetic power spectra as functions of wavenumber and (d) the ratio between the two spectra from the simulation at the end of the simulation $t = 1600\,\omega_{ci}^{-1}$.}
    \label{fig:comparison}
\end{figure}

Figure \ref{fig:comparison} compares the electromagnetic power spectra from MMS observations and simulations. Both observed and simulated spectra transition from an electromagnetic regime with $c \vert \delta E \vert / v_\mathrm{A} \vert \delta B \vert \sim 1$ in the low-frequency ($\omega / \omega_{ci} < 1$) or long-wavelength ($k d_i < 1$) limit, to an electrostatic one with $c \vert \delta E \vert / v_\mathrm{A} \vert \delta B \vert \gg 1$ in the high-frequency ($\omega \lesssim \omega_{pi}$) or short-wavelength ($k \lambda_D \lesssim 1$) limit.

In the electromagnetic regime, the dominant Alfv\'en wave at $0.07 \omega_{ci}$ from the MMS observation generates second and third harmonics at $\omega / \omega_{ci} = 0.14, 0.28$ due to phase steepening, similar to the simulation. The simulation shows that long-wavelength electrostatic field energy can be comparable to transverse electric field energy [Figures \ref{fig:hybrid-propagation}(a) \ref{fig:hybrid-propagation}(c), and \ref{fig:comparison}(a)--(b)], with the ordering $\langle \delta E_x \rangle^2 \sim \delta E_\perp^2 \sim 0.01 (v_\mathrm{A} B_0 /c)^2 \sim (v_\mathrm{A} \delta B_\perp /c)^2 \ll (\delta B_\perp)^2$. However, these relatively small-amplitude, long-wavelength electrostatic fields mediate energy transfer from fluid-scale Alfv\'en waves to kinetic-scale IAWs and core ion heating.

In the electrostatic regime, the Debye-scale IAWs exhibit a relatively flat electric field spectrum compared to the magnetic field spectrum. In the hybrid-kinetic simulation, electrostatic energy accumulates near the grid scale due to the absence of an intrinsic Debye scale to terminate the energy transfer. Nevertheless, the transition from electromagnetic fluctuations at the fluid scale to electrostatic fluctuations at the kinetic scale is clearly demonstrated in both the observations and simulations.

In summary, we provide direct observational and computational evidence supporting electrostatic-driven energy transfer from fluid-scale Alfv\'en waves to kinetic-scale IAWs. In a realistic parameter regime of $v_\mathrm{Ti} \sim c_s < v_\mathrm{A}$, two ion beams centered at $c_s$ and $v_\mathrm{A}$ are simultaneously accelerated through nonlinear Landau resonance by two coexisting compressive electrostatic modes, both modes driven by phase-steepened Alfv\'en waves. This process results in the generation of kinetic-scale IAWs, substantial heating of thermal ions, and acceleration of suprathermal ions up to $v_\mathrm{A}$. More broadly, the Alfv\'en-acoustic channelling of ion energy may contribute to ion heating near various plasma boundaries in the interplanetary space, such as magnetic discontinuities \cite{woodham2021enhanced,malaspina2023statistical}, and low-velocity regions ahead of high-speed streams \cite{gurnett1979ion}. Moreover, given that the wavelengths of kinetic-scale IAWs are comparable to the electron thermal gyroradius \cite{kamaletdinov2022quantifying,kamaletdinov2024nonlinear}, the Alfv\'en-acoustic channelling may also contribute to the momentum exchange between ions and electrons and electron heating \cite{mozer2022core}, thereby providing a collisionless dissipation mechanism to allow fast magnetic reconnection in current sheets \cite{coroniti1977magnetic,sagdeev19791976,smith1972current,coppi1971processes}.

\begin{acknowledgments}
This work was supported by NASA grants NO.~80NSSC22K1634 and No.~80NSSC23K0086 and NSF grant NO.~2108582. We acknowledge MMS data (including FGM, EDP, and FPI) obtained from \url{https://lasp.colorado.edu/mms/sdc/public/}. Data access and processing was done using SPEDAS V4.1 \cite{Angelopoulos19}. We would like to acknowledge high-performance computing support from Derecho (\url{https://doi.org/10.5065/qx9a-pg09}) provided by NCAR's Computational and Information Systems Laboratory, sponsored by the National Science Foundation \cite{derecho}. We wish to thank Marco Velli for helpful discussions.
\end{acknowledgments}





\begin{thebibliography}{68}%
\makeatletter
\providecommand \@ifxundefined [1]{%
 \@ifx{#1\undefined}
}%
\providecommand \@ifnum [1]{%
 \ifnum #1\expandafter \@firstoftwo
 \else \expandafter \@secondoftwo
 \fi
}%
\providecommand \@ifx [1]{%
 \ifx #1\expandafter \@firstoftwo
 \else \expandafter \@secondoftwo
 \fi
}%
\providecommand \natexlab [1]{#1}%
\providecommand \enquote  [1]{``#1''}%
\providecommand \bibnamefont  [1]{#1}%
\providecommand \bibfnamefont [1]{#1}%
\providecommand \citenamefont [1]{#1}%
\providecommand \href@noop [0]{\@secondoftwo}%
\providecommand \href [0]{\begingroup \@sanitize@url \@href}%
\providecommand \@href[1]{\@@startlink{#1}\@@href}%
\providecommand \@@href[1]{\endgroup#1\@@endlink}%
\providecommand \@sanitize@url [0]{\catcode `\\12\catcode `\$12\catcode `\&12\catcode `\#12\catcode `\^12\catcode `\_12\catcode `\%12\relax}%
\providecommand \@@startlink[1]{}%
\providecommand \@@endlink[0]{}%
\providecommand \url  [0]{\begingroup\@sanitize@url \@url }%
\providecommand \@url [1]{\endgroup\@href {#1}{\urlprefix }}%
\providecommand \urlprefix  [0]{URL }%
\providecommand \Eprint [0]{\href }%
\providecommand \doibase [0]{http://dx.doi.org/}%
\providecommand \selectlanguage [0]{\@gobble}%
\providecommand \bibinfo  [0]{\@secondoftwo}%
\providecommand \bibfield  [0]{\@secondoftwo}%
\providecommand \translation [1]{[#1]}%
\providecommand \BibitemOpen [0]{}%
\providecommand \bibitemStop [0]{}%
\providecommand \bibitemNoStop [0]{.\EOS\space}%
\providecommand \EOS [0]{\spacefactor3000\relax}%
\providecommand \BibitemShut  [1]{\csname bibitem#1\endcsname}%
\let\auto@bib@innerbib\@empty
\bibitem [{\citenamefont {{Iroshnikov}}(1964)}]{Iroshnikov64}%
  \BibitemOpen
  \bibfield  {author} {\bibinfo {author} {\bibfnamefont {P.~S.}\ \bibnamefont {{Iroshnikov}}},\ }\href@noop {} {\bibfield  {journal} {\bibinfo  {journal} {Soviet Astronomy}\ }\textbf {\bibinfo {volume} {7}},\ \bibinfo {pages} {566} (\bibinfo {year} {1964})}\BibitemShut {NoStop}%
\bibitem [{\citenamefont {{Kraichnan}}(1965)}]{Kraichnan65}%
  \BibitemOpen
  \bibfield  {author} {\bibinfo {author} {\bibfnamefont {R.~H.}\ \bibnamefont {{Kraichnan}}},\ }\href {\doibase 10.1063/1.1761412} {\bibfield  {journal} {\bibinfo  {journal} {Physics of Fluids}\ }\textbf {\bibinfo {volume} {8}},\ \bibinfo {pages} {1385} (\bibinfo {year} {1965})}\BibitemShut {NoStop}%
\bibitem [{\citenamefont {{Goldreich}}\ and\ \citenamefont {{Sridhar}}(1997)}]{Goldreich&Sridhar97}%
  \BibitemOpen
  \bibfield  {author} {\bibinfo {author} {\bibfnamefont {P.}~\bibnamefont {{Goldreich}}}\ and\ \bibinfo {author} {\bibfnamefont {S.}~\bibnamefont {{Sridhar}}},\ }\href@noop {} {\bibfield  {journal} {\bibinfo  {journal} {\apj}\ }\textbf {\bibinfo {volume} {485}},\ \bibinfo {pages} {680} (\bibinfo {year} {1997})},\ \Eprint {http://arxiv.org/abs/astro-ph/9612243} {astro-ph/9612243} \BibitemShut {NoStop}%
\bibitem [{\citenamefont {Boldyrev}(2006)}]{boldyrev2006spectrum}%
  \BibitemOpen
  \bibfield  {author} {\bibinfo {author} {\bibfnamefont {S.}~\bibnamefont {Boldyrev}},\ }\href {\doibase 10.1103/PhysRevLett.96.115002} {\bibfield  {journal} {\bibinfo  {journal} {Phys. Rev. Lett.}\ }\textbf {\bibinfo {volume} {96}},\ \bibinfo {pages} {115002} (\bibinfo {year} {2006})}\BibitemShut {NoStop}%
\bibitem [{\citenamefont {Howes}\ \emph {et~al.}(2012)\citenamefont {Howes}, \citenamefont {Drake}, \citenamefont {Nielson}, \citenamefont {Carter}, \citenamefont {Kletzing},\ and\ \citenamefont {Skiff}}]{howes2012toward}%
  \BibitemOpen
  \bibfield  {author} {\bibinfo {author} {\bibfnamefont {G.}~\bibnamefont {Howes}}, \bibinfo {author} {\bibfnamefont {D.}~\bibnamefont {Drake}}, \bibinfo {author} {\bibfnamefont {K.}~\bibnamefont {Nielson}}, \bibinfo {author} {\bibfnamefont {T.}~\bibnamefont {Carter}}, \bibinfo {author} {\bibfnamefont {C.}~\bibnamefont {Kletzing}}, \ and\ \bibinfo {author} {\bibfnamefont {F.}~\bibnamefont {Skiff}},\ }\href@noop {} {\bibfield  {journal} {\bibinfo  {journal} {Physical review letters}\ }\textbf {\bibinfo {volume} {109}},\ \bibinfo {pages} {255001} (\bibinfo {year} {2012})}\BibitemShut {NoStop}%
\bibitem [{\citenamefont {Hollweg}(1971)}]{hollweg1971density}%
  \BibitemOpen
  \bibfield  {author} {\bibinfo {author} {\bibfnamefont {J.~V.}\ \bibnamefont {Hollweg}},\ }\href@noop {} {\bibfield  {journal} {\bibinfo  {journal} {Journal of Geophysical Research}\ }\textbf {\bibinfo {volume} {76}},\ \bibinfo {pages} {5155} (\bibinfo {year} {1971})}\BibitemShut {NoStop}%
\bibitem [{\citenamefont {Medvedev}\ \emph {et~al.}(1998)\citenamefont {Medvedev}, \citenamefont {Diamond}, \citenamefont {Rosenbluth},\ and\ \citenamefont {Shevchenko}}]{medvedev1998asymptotic}%
  \BibitemOpen
  \bibfield  {author} {\bibinfo {author} {\bibfnamefont {M.}~\bibnamefont {Medvedev}}, \bibinfo {author} {\bibfnamefont {P.}~\bibnamefont {Diamond}}, \bibinfo {author} {\bibfnamefont {M.}~\bibnamefont {Rosenbluth}}, \ and\ \bibinfo {author} {\bibfnamefont {V.}~\bibnamefont {Shevchenko}},\ }\href@noop {} {\bibfield  {journal} {\bibinfo  {journal} {Physical review letters}\ }\textbf {\bibinfo {volume} {81}},\ \bibinfo {pages} {5824} (\bibinfo {year} {1998})}\BibitemShut {NoStop}%
\bibitem [{\citenamefont {Araneda}\ \emph {et~al.}(2008)\citenamefont {Araneda}, \citenamefont {Marsch}, \citenamefont {Adolfo} \emph {et~al.}}]{araneda2008proton}%
  \BibitemOpen
  \bibfield  {author} {\bibinfo {author} {\bibfnamefont {J.~A.}\ \bibnamefont {Araneda}}, \bibinfo {author} {\bibfnamefont {E.}~\bibnamefont {Marsch}}, \bibinfo {author} {\bibfnamefont {F.}~\bibnamefont {Adolfo}},  \emph {et~al.},\ }\href@noop {} {\bibfield  {journal} {\bibinfo  {journal} {Physical review letters}\ }\textbf {\bibinfo {volume} {100}},\ \bibinfo {pages} {125003} (\bibinfo {year} {2008})}\BibitemShut {NoStop}%
\bibitem [{\citenamefont {Matteini}\ \emph {et~al.}(2010)\citenamefont {Matteini}, \citenamefont {Landi}, \citenamefont {Velli},\ and\ \citenamefont {Hellinger}}]{matteini2010kinetics}%
  \BibitemOpen
  \bibfield  {author} {\bibinfo {author} {\bibfnamefont {L.}~\bibnamefont {Matteini}}, \bibinfo {author} {\bibfnamefont {S.}~\bibnamefont {Landi}}, \bibinfo {author} {\bibfnamefont {M.}~\bibnamefont {Velli}}, \ and\ \bibinfo {author} {\bibfnamefont {P.}~\bibnamefont {Hellinger}},\ }\href@noop {} {\bibfield  {journal} {\bibinfo  {journal} {Journal of Geophysical Research: Space Physics}\ }\textbf {\bibinfo {volume} {115}} (\bibinfo {year} {2010})}\BibitemShut {NoStop}%
\bibitem [{\citenamefont {Valentini}\ \emph {et~al.}(2008)\citenamefont {Valentini}, \citenamefont {Veltri}, \citenamefont {Califano},\ and\ \citenamefont {Mangeney}}]{valentini2008cross}%
  \BibitemOpen
  \bibfield  {author} {\bibinfo {author} {\bibfnamefont {F.}~\bibnamefont {Valentini}}, \bibinfo {author} {\bibfnamefont {P.}~\bibnamefont {Veltri}}, \bibinfo {author} {\bibfnamefont {F.}~\bibnamefont {Califano}}, \ and\ \bibinfo {author} {\bibfnamefont {A.}~\bibnamefont {Mangeney}},\ }\href@noop {} {\bibfield  {journal} {\bibinfo  {journal} {Physical review letters}\ }\textbf {\bibinfo {volume} {101}},\ \bibinfo {pages} {025006} (\bibinfo {year} {2008})}\BibitemShut {NoStop}%
\bibitem [{\citenamefont {Valentini}\ and\ \citenamefont {Veltri}(2009)}]{valentini2009electrostatic}%
  \BibitemOpen
  \bibfield  {author} {\bibinfo {author} {\bibfnamefont {F.}~\bibnamefont {Valentini}}\ and\ \bibinfo {author} {\bibfnamefont {P.}~\bibnamefont {Veltri}},\ }\href@noop {} {\bibfield  {journal} {\bibinfo  {journal} {Physical review letters}\ }\textbf {\bibinfo {volume} {102}},\ \bibinfo {pages} {225001} (\bibinfo {year} {2009})}\BibitemShut {NoStop}%
\bibitem [{\citenamefont {Valentini}\ \emph {et~al.}(2011)\citenamefont {Valentini}, \citenamefont {Perrone},\ and\ \citenamefont {Veltri}}]{valentini2011short}%
  \BibitemOpen
  \bibfield  {author} {\bibinfo {author} {\bibfnamefont {F.}~\bibnamefont {Valentini}}, \bibinfo {author} {\bibfnamefont {D.}~\bibnamefont {Perrone}}, \ and\ \bibinfo {author} {\bibfnamefont {P.}~\bibnamefont {Veltri}},\ }\href@noop {} {\bibfield  {journal} {\bibinfo  {journal} {The Astrophysical Journal}\ }\textbf {\bibinfo {volume} {739}},\ \bibinfo {pages} {54} (\bibinfo {year} {2011})}\BibitemShut {NoStop}%
\bibitem [{\citenamefont {Valentini}\ \emph {et~al.}(2014)\citenamefont {Valentini}, \citenamefont {Vecchio}, \citenamefont {Donato}, \citenamefont {Carbone}, \citenamefont {Briand}, \citenamefont {Bougeret},\ and\ \citenamefont {Veltri}}]{valentini2014nonlinear}%
  \BibitemOpen
  \bibfield  {author} {\bibinfo {author} {\bibfnamefont {F.}~\bibnamefont {Valentini}}, \bibinfo {author} {\bibfnamefont {A.}~\bibnamefont {Vecchio}}, \bibinfo {author} {\bibfnamefont {S.}~\bibnamefont {Donato}}, \bibinfo {author} {\bibfnamefont {V.}~\bibnamefont {Carbone}}, \bibinfo {author} {\bibfnamefont {C.}~\bibnamefont {Briand}}, \bibinfo {author} {\bibfnamefont {J.}~\bibnamefont {Bougeret}}, \ and\ \bibinfo {author} {\bibfnamefont {P.}~\bibnamefont {Veltri}},\ }\href@noop {} {\bibfield  {journal} {\bibinfo  {journal} {The Astrophysical Journal Letters}\ }\textbf {\bibinfo {volume} {788}},\ \bibinfo {pages} {L16} (\bibinfo {year} {2014})}\BibitemShut {NoStop}%
\bibitem [{\citenamefont {Graham}\ \emph {et~al.}(2021)\citenamefont {Graham}, \citenamefont {Khotyaintsev}, \citenamefont {Vaivads}, \citenamefont {Edberg}, \citenamefont {Eriksson}, \citenamefont {Johansson}, \citenamefont {Sorriso-Valvo}, \citenamefont {Maksimovic}, \citenamefont {Sou{\v{c}}ek}, \citenamefont {P{\'\i}{\v{s}}a} \emph {et~al.}}]{graham2021kinetic}%
  \BibitemOpen
  \bibfield  {author} {\bibinfo {author} {\bibfnamefont {D.~B.}\ \bibnamefont {Graham}}, \bibinfo {author} {\bibfnamefont {Y.~V.}\ \bibnamefont {Khotyaintsev}}, \bibinfo {author} {\bibfnamefont {A.}~\bibnamefont {Vaivads}}, \bibinfo {author} {\bibfnamefont {N.~J.}\ \bibnamefont {Edberg}}, \bibinfo {author} {\bibfnamefont {A.~I.}\ \bibnamefont {Eriksson}}, \bibinfo {author} {\bibfnamefont {E.~P.}\ \bibnamefont {Johansson}}, \bibinfo {author} {\bibfnamefont {L.}~\bibnamefont {Sorriso-Valvo}}, \bibinfo {author} {\bibfnamefont {M.}~\bibnamefont {Maksimovic}}, \bibinfo {author} {\bibfnamefont {J.}~\bibnamefont {Sou{\v{c}}ek}}, \bibinfo {author} {\bibfnamefont {D.}~\bibnamefont {P{\'\i}{\v{s}}a}},  \emph {et~al.},\ }\href@noop {} {\bibfield  {journal} {\bibinfo  {journal} {Astronomy \& Astrophysics}\ }\textbf {\bibinfo {volume} {656}},\ \bibinfo {pages} {A23} (\bibinfo {year} {2021})}\BibitemShut {NoStop}%
\bibitem [{\citenamefont {Mozer}\ \emph {et~al.}(2020)\citenamefont {Mozer}, \citenamefont {Bonnell}, \citenamefont {Bowen}, \citenamefont {Schumm},\ and\ \citenamefont {Vasko}}]{mozer2020large}%
  \BibitemOpen
  \bibfield  {author} {\bibinfo {author} {\bibfnamefont {F.}~\bibnamefont {Mozer}}, \bibinfo {author} {\bibfnamefont {J.}~\bibnamefont {Bonnell}}, \bibinfo {author} {\bibfnamefont {T.}~\bibnamefont {Bowen}}, \bibinfo {author} {\bibfnamefont {G.}~\bibnamefont {Schumm}}, \ and\ \bibinfo {author} {\bibfnamefont {I.}~\bibnamefont {Vasko}},\ }\href@noop {} {\bibfield  {journal} {\bibinfo  {journal} {The Astrophysical Journal}\ }\textbf {\bibinfo {volume} {901}},\ \bibinfo {pages} {107} (\bibinfo {year} {2020})}\BibitemShut {NoStop}%
\bibitem [{\citenamefont {Malaspina}\ \emph {et~al.}(2023)\citenamefont {Malaspina}, \citenamefont {Kromyda}, \citenamefont {Ergun},\ and\ \citenamefont {Livi}}]{malaspina2023statistical}%
  \BibitemOpen
  \bibfield  {author} {\bibinfo {author} {\bibfnamefont {D.}~\bibnamefont {Malaspina}}, \bibinfo {author} {\bibfnamefont {L.}~\bibnamefont {Kromyda}}, \bibinfo {author} {\bibfnamefont {R.}~\bibnamefont {Ergun}}, \ and\ \bibinfo {author} {\bibfnamefont {R.}~\bibnamefont {Livi}},\ }\href@noop {} {\bibfield  {journal} {\bibinfo  {journal} {AGU23}\ } (\bibinfo {year} {2023})}\BibitemShut {NoStop}%
\bibitem [{\citenamefont {{Malaspina}}\ \emph {et~al.}(2024)\citenamefont {{Malaspina}}, \citenamefont {{Ergun}}, \citenamefont {{Cairns}}, \citenamefont {{Short}}, \citenamefont {{Verniero}}, \citenamefont {{Cattell}},\ and\ \citenamefont {{Livi}}}]{malaspina2024frequency}%
  \BibitemOpen
  \bibfield  {author} {\bibinfo {author} {\bibfnamefont {D.~M.}\ \bibnamefont {{Malaspina}}}, \bibinfo {author} {\bibfnamefont {R.~E.}\ \bibnamefont {{Ergun}}}, \bibinfo {author} {\bibfnamefont {I.~H.}\ \bibnamefont {{Cairns}}}, \bibinfo {author} {\bibfnamefont {B.}~\bibnamefont {{Short}}}, \bibinfo {author} {\bibfnamefont {J.~L.}\ \bibnamefont {{Verniero}}}, \bibinfo {author} {\bibfnamefont {C.}~\bibnamefont {{Cattell}}}, \ and\ \bibinfo {author} {\bibfnamefont {R.}~\bibnamefont {{Livi}}},\ }\href {\doibase 10.3847/1538-4357/ad4b12} {\bibfield  {journal} {\bibinfo  {journal} {\apj}\ }\textbf {\bibinfo {volume} {969}},\ \bibinfo {eid} {60} (\bibinfo {year} {2024})}\BibitemShut {NoStop}%
\bibitem [{\citenamefont {Squire}\ \emph {et~al.}(2020)\citenamefont {Squire}, \citenamefont {Chandran},\ and\ \citenamefont {Meyrand}}]{squire2020situ}%
  \BibitemOpen
  \bibfield  {author} {\bibinfo {author} {\bibfnamefont {J.}~\bibnamefont {Squire}}, \bibinfo {author} {\bibfnamefont {B.~D.}\ \bibnamefont {Chandran}}, \ and\ \bibinfo {author} {\bibfnamefont {R.}~\bibnamefont {Meyrand}},\ }\href@noop {} {\bibfield  {journal} {\bibinfo  {journal} {The Astrophysical Journal Letters}\ }\textbf {\bibinfo {volume} {891}},\ \bibinfo {pages} {L2} (\bibinfo {year} {2020})}\BibitemShut {NoStop}%
\bibitem [{\citenamefont {Mallet}\ \emph {et~al.}(2021)\citenamefont {Mallet}, \citenamefont {Squire}, \citenamefont {Chandran}, \citenamefont {Bowen},\ and\ \citenamefont {Bale}}]{mallet2021evolution}%
  \BibitemOpen
  \bibfield  {author} {\bibinfo {author} {\bibfnamefont {A.}~\bibnamefont {Mallet}}, \bibinfo {author} {\bibfnamefont {J.}~\bibnamefont {Squire}}, \bibinfo {author} {\bibfnamefont {B.~D.}\ \bibnamefont {Chandran}}, \bibinfo {author} {\bibfnamefont {T.}~\bibnamefont {Bowen}}, \ and\ \bibinfo {author} {\bibfnamefont {S.~D.}\ \bibnamefont {Bale}},\ }\href@noop {} {\bibfield  {journal} {\bibinfo  {journal} {The Astrophysical Journal}\ }\textbf {\bibinfo {volume} {918}},\ \bibinfo {pages} {62} (\bibinfo {year} {2021})}\BibitemShut {NoStop}%
\bibitem [{\citenamefont {Tenerani}\ \emph {et~al.}(2021)\citenamefont {Tenerani}, \citenamefont {Sioulas}, \citenamefont {Matteini}, \citenamefont {Panasenco}, \citenamefont {Shi},\ and\ \citenamefont {Velli}}]{tenerani2021evolution}%
  \BibitemOpen
  \bibfield  {author} {\bibinfo {author} {\bibfnamefont {A.}~\bibnamefont {Tenerani}}, \bibinfo {author} {\bibfnamefont {N.}~\bibnamefont {Sioulas}}, \bibinfo {author} {\bibfnamefont {L.}~\bibnamefont {Matteini}}, \bibinfo {author} {\bibfnamefont {O.}~\bibnamefont {Panasenco}}, \bibinfo {author} {\bibfnamefont {C.}~\bibnamefont {Shi}}, \ and\ \bibinfo {author} {\bibfnamefont {M.}~\bibnamefont {Velli}},\ }\href@noop {} {\bibfield  {journal} {\bibinfo  {journal} {The Astrophysical Journal Letters}\ }\textbf {\bibinfo {volume} {919}},\ \bibinfo {pages} {L31} (\bibinfo {year} {2021})}\BibitemShut {NoStop}%
\bibitem [{\citenamefont {Mozer}\ \emph {et~al.}(2022)\citenamefont {Mozer}, \citenamefont {Bale}, \citenamefont {Cattell}, \citenamefont {Halekas}, \citenamefont {Vasko}, \citenamefont {Verniero},\ and\ \citenamefont {Kellogg}}]{mozer2022core}%
  \BibitemOpen
  \bibfield  {author} {\bibinfo {author} {\bibfnamefont {F.}~\bibnamefont {Mozer}}, \bibinfo {author} {\bibfnamefont {S.}~\bibnamefont {Bale}}, \bibinfo {author} {\bibfnamefont {C.}~\bibnamefont {Cattell}}, \bibinfo {author} {\bibfnamefont {J.}~\bibnamefont {Halekas}}, \bibinfo {author} {\bibfnamefont {I.}~\bibnamefont {Vasko}}, \bibinfo {author} {\bibfnamefont {J.}~\bibnamefont {Verniero}}, \ and\ \bibinfo {author} {\bibfnamefont {P.}~\bibnamefont {Kellogg}},\ }\href@noop {} {\bibfield  {journal} {\bibinfo  {journal} {The Astrophysical Journal Letters}\ }\textbf {\bibinfo {volume} {927}},\ \bibinfo {pages} {L15} (\bibinfo {year} {2022})}\BibitemShut {NoStop}%
\bibitem [{\citenamefont {Kellogg}\ \emph {et~al.}(2024)\citenamefont {Kellogg}, \citenamefont {Mozer}, \citenamefont {Moncuquet}, \citenamefont {Malaspina}, \citenamefont {Halekas}, \citenamefont {Bale},\ and\ \citenamefont {Goetz}}]{kellogg2024heating}%
  \BibitemOpen
  \bibfield  {author} {\bibinfo {author} {\bibfnamefont {P.}~\bibnamefont {Kellogg}}, \bibinfo {author} {\bibfnamefont {F.}~\bibnamefont {Mozer}}, \bibinfo {author} {\bibfnamefont {M.}~\bibnamefont {Moncuquet}}, \bibinfo {author} {\bibfnamefont {D.}~\bibnamefont {Malaspina}}, \bibinfo {author} {\bibfnamefont {J.}~\bibnamefont {Halekas}}, \bibinfo {author} {\bibfnamefont {S.}~\bibnamefont {Bale}}, \ and\ \bibinfo {author} {\bibfnamefont {K.}~\bibnamefont {Goetz}},\ }\href@noop {} {\bibfield  {journal} {\bibinfo  {journal} {The Astrophysical Journal}\ }\textbf {\bibinfo {volume} {964}},\ \bibinfo {pages} {68} (\bibinfo {year} {2024})}\BibitemShut {NoStop}%
\bibitem [{\citenamefont {Woodham}\ \emph {et~al.}(2021)\citenamefont {Woodham}, \citenamefont {Horbury}, \citenamefont {Matteini}, \citenamefont {Woolley}, \citenamefont {Laker}, \citenamefont {Bale}, \citenamefont {Nicolaou}, \citenamefont {Stawarz}, \citenamefont {Stansby}, \citenamefont {Hietala} \emph {et~al.}}]{woodham2021enhanced}%
  \BibitemOpen
  \bibfield  {author} {\bibinfo {author} {\bibfnamefont {L.}~\bibnamefont {Woodham}}, \bibinfo {author} {\bibfnamefont {T.}~\bibnamefont {Horbury}}, \bibinfo {author} {\bibfnamefont {L.}~\bibnamefont {Matteini}}, \bibinfo {author} {\bibfnamefont {T.}~\bibnamefont {Woolley}}, \bibinfo {author} {\bibfnamefont {R.}~\bibnamefont {Laker}}, \bibinfo {author} {\bibfnamefont {S.}~\bibnamefont {Bale}}, \bibinfo {author} {\bibfnamefont {G.}~\bibnamefont {Nicolaou}}, \bibinfo {author} {\bibfnamefont {J.}~\bibnamefont {Stawarz}}, \bibinfo {author} {\bibfnamefont {D.}~\bibnamefont {Stansby}}, \bibinfo {author} {\bibfnamefont {H.}~\bibnamefont {Hietala}},  \emph {et~al.},\ }\href@noop {} {\bibfield  {journal} {\bibinfo  {journal} {Astronomy \& Astrophysics}\ }\textbf {\bibinfo {volume} {650}},\ \bibinfo {pages} {L1} (\bibinfo {year} {2021})}\BibitemShut {NoStop}%
\bibitem [{\citenamefont {Gorelenkov}\ \emph {et~al.}(2009)\citenamefont {Gorelenkov}, \citenamefont {Van~Zeeland}, \citenamefont {Berk}, \citenamefont {Crocker}, \citenamefont {Darrow}, \citenamefont {Fredrickson}, \citenamefont {Fu}, \citenamefont {Heidbrink}, \citenamefont {Menard},\ and\ \citenamefont {Nazikian}}]{gorelenkov2009beta}%
  \BibitemOpen
  \bibfield  {author} {\bibinfo {author} {\bibfnamefont {N.}~\bibnamefont {Gorelenkov}}, \bibinfo {author} {\bibfnamefont {M.}~\bibnamefont {Van~Zeeland}}, \bibinfo {author} {\bibfnamefont {H.}~\bibnamefont {Berk}}, \bibinfo {author} {\bibfnamefont {N.}~\bibnamefont {Crocker}}, \bibinfo {author} {\bibfnamefont {D.}~\bibnamefont {Darrow}}, \bibinfo {author} {\bibfnamefont {E.}~\bibnamefont {Fredrickson}}, \bibinfo {author} {\bibfnamefont {G.-Y.}\ \bibnamefont {Fu}}, \bibinfo {author} {\bibfnamefont {W.}~\bibnamefont {Heidbrink}}, \bibinfo {author} {\bibfnamefont {J.}~\bibnamefont {Menard}}, \ and\ \bibinfo {author} {\bibfnamefont {R.}~\bibnamefont {Nazikian}},\ }\href@noop {} {\bibfield  {journal} {\bibinfo  {journal} {Physics of Plasmas}\ }\textbf {\bibinfo {volume} {16}} (\bibinfo {year} {2009})}\BibitemShut {NoStop}%
\bibitem [{\citenamefont {Curran}\ \emph {et~al.}(2012)\citenamefont {Curran}, \citenamefont {Lauber}, \citenamefont {Mc~Carthy}, \citenamefont {da~Graca}, \citenamefont {Igochine}, \citenamefont {Team} \emph {et~al.}}]{curran2012low}%
  \BibitemOpen
  \bibfield  {author} {\bibinfo {author} {\bibfnamefont {D.}~\bibnamefont {Curran}}, \bibinfo {author} {\bibfnamefont {P.}~\bibnamefont {Lauber}}, \bibinfo {author} {\bibfnamefont {P.}~\bibnamefont {Mc~Carthy}}, \bibinfo {author} {\bibfnamefont {S.}~\bibnamefont {da~Graca}}, \bibinfo {author} {\bibfnamefont {V.}~\bibnamefont {Igochine}}, \bibinfo {author} {\bibfnamefont {A.~U.}\ \bibnamefont {Team}},  \emph {et~al.},\ }\href@noop {} {\bibfield  {journal} {\bibinfo  {journal} {Plasma Physics and Controlled Fusion}\ }\textbf {\bibinfo {volume} {54}},\ \bibinfo {pages} {055001} (\bibinfo {year} {2012})}\BibitemShut {NoStop}%
\bibitem [{\citenamefont {Bierwage}\ \emph {et~al.}(2015)\citenamefont {Bierwage}, \citenamefont {Aiba},\ and\ \citenamefont {Shinohara}}]{bierwage2015alfven}%
  \BibitemOpen
  \bibfield  {author} {\bibinfo {author} {\bibfnamefont {A.}~\bibnamefont {Bierwage}}, \bibinfo {author} {\bibfnamefont {N.}~\bibnamefont {Aiba}}, \ and\ \bibinfo {author} {\bibfnamefont {K.}~\bibnamefont {Shinohara}},\ }\href@noop {} {\bibfield  {journal} {\bibinfo  {journal} {Physical Review Letters}\ }\textbf {\bibinfo {volume} {114}},\ \bibinfo {pages} {015002} (\bibinfo {year} {2015})}\BibitemShut {NoStop}%
\bibitem [{\citenamefont {Chen}\ and\ \citenamefont {Zonca}(2016)}]{chen2016physics}%
  \BibitemOpen
  \bibfield  {author} {\bibinfo {author} {\bibfnamefont {L.}~\bibnamefont {Chen}}\ and\ \bibinfo {author} {\bibfnamefont {F.}~\bibnamefont {Zonca}},\ }\href@noop {} {\bibfield  {journal} {\bibinfo  {journal} {Reviews of Modern Physics}\ }\textbf {\bibinfo {volume} {88}},\ \bibinfo {pages} {015008} (\bibinfo {year} {2016})}\BibitemShut {NoStop}%
\bibitem [{\citenamefont {{Burch}}\ \emph {et~al.}(2016)\citenamefont {{Burch}}, \citenamefont {{Moore}}, \citenamefont {{Torbert}},\ and\ \citenamefont {{Giles}}}]{Burch16}%
  \BibitemOpen
  \bibfield  {author} {\bibinfo {author} {\bibfnamefont {J.~L.}\ \bibnamefont {{Burch}}}, \bibinfo {author} {\bibfnamefont {T.~E.}\ \bibnamefont {{Moore}}}, \bibinfo {author} {\bibfnamefont {R.~B.}\ \bibnamefont {{Torbert}}}, \ and\ \bibinfo {author} {\bibfnamefont {B.~L.}\ \bibnamefont {{Giles}}},\ }\href {\doibase 10.1007/s11214-015-0164-9} {\bibfield  {journal} {\bibinfo  {journal} {\ssr}\ }\textbf {\bibinfo {volume} {199}},\ \bibinfo {pages} {5} (\bibinfo {year} {2016})}\BibitemShut {NoStop}%
\bibitem [{\citenamefont {Eriksson}\ \emph {et~al.}(2016)\citenamefont {Eriksson}, \citenamefont {Lavraud}, \citenamefont {Wilder}, \citenamefont {Stawarz}, \citenamefont {Giles}, \citenamefont {Burch}, \citenamefont {Baumjohann}, \citenamefont {Ergun}, \citenamefont {Lindqvist}, \citenamefont {Magnes} \emph {et~al.}}]{eriksson2016magnetospheric}%
  \BibitemOpen
  \bibfield  {author} {\bibinfo {author} {\bibfnamefont {S.}~\bibnamefont {Eriksson}}, \bibinfo {author} {\bibfnamefont {B.}~\bibnamefont {Lavraud}}, \bibinfo {author} {\bibfnamefont {F.}~\bibnamefont {Wilder}}, \bibinfo {author} {\bibfnamefont {J.}~\bibnamefont {Stawarz}}, \bibinfo {author} {\bibfnamefont {B.}~\bibnamefont {Giles}}, \bibinfo {author} {\bibfnamefont {J.}~\bibnamefont {Burch}}, \bibinfo {author} {\bibfnamefont {W.}~\bibnamefont {Baumjohann}}, \bibinfo {author} {\bibfnamefont {R.}~\bibnamefont {Ergun}}, \bibinfo {author} {\bibfnamefont {P.-A.}\ \bibnamefont {Lindqvist}}, \bibinfo {author} {\bibfnamefont {W.}~\bibnamefont {Magnes}},  \emph {et~al.},\ }\href@noop {} {\bibfield  {journal} {\bibinfo  {journal} {Geophysical Research Letters}\ }\textbf {\bibinfo {volume} {43}},\ \bibinfo {pages} {5606} (\bibinfo {year} {2016})}\BibitemShut {NoStop}%
\bibitem [{\citenamefont {Stawarz}\ \emph {et~al.}(2016)\citenamefont {Stawarz}, \citenamefont {Eriksson}, \citenamefont {Wilder}, \citenamefont {Ergun}, \citenamefont {Schwartz}, \citenamefont {Pouquet}, \citenamefont {Burch}, \citenamefont {Giles}, \citenamefont {Khotyaintsev}, \citenamefont {Contel} \emph {et~al.}}]{stawarz2016observations}%
  \BibitemOpen
  \bibfield  {author} {\bibinfo {author} {\bibfnamefont {J.}~\bibnamefont {Stawarz}}, \bibinfo {author} {\bibfnamefont {S.}~\bibnamefont {Eriksson}}, \bibinfo {author} {\bibfnamefont {F.}~\bibnamefont {Wilder}}, \bibinfo {author} {\bibfnamefont {R.}~\bibnamefont {Ergun}}, \bibinfo {author} {\bibfnamefont {S.}~\bibnamefont {Schwartz}}, \bibinfo {author} {\bibfnamefont {A.}~\bibnamefont {Pouquet}}, \bibinfo {author} {\bibfnamefont {J.}~\bibnamefont {Burch}}, \bibinfo {author} {\bibfnamefont {B.}~\bibnamefont {Giles}}, \bibinfo {author} {\bibfnamefont {Y.}~\bibnamefont {Khotyaintsev}}, \bibinfo {author} {\bibfnamefont {O.~L.}\ \bibnamefont {Contel}},  \emph {et~al.},\ }\href@noop {} {\bibfield  {journal} {\bibinfo  {journal} {Journal of Geophysical Research: Space Physics}\ }\textbf {\bibinfo {volume} {121}},\ \bibinfo {pages} {11} (\bibinfo {year} {2016})}\BibitemShut {NoStop}%
\bibitem [{\citenamefont {Sorriso-Valvo}\ \emph {et~al.}(2019)\citenamefont {Sorriso-Valvo}, \citenamefont {Catapano}, \citenamefont {Retin{\`o}}, \citenamefont {Le~Contel}, \citenamefont {Perrone}, \citenamefont {Roberts}, \citenamefont {Coburn}, \citenamefont {Panebianco}, \citenamefont {Valentini}, \citenamefont {Perri} \emph {et~al.}}]{sorriso2019turbulence}%
  \BibitemOpen
  \bibfield  {author} {\bibinfo {author} {\bibfnamefont {L.}~\bibnamefont {Sorriso-Valvo}}, \bibinfo {author} {\bibfnamefont {F.}~\bibnamefont {Catapano}}, \bibinfo {author} {\bibfnamefont {A.}~\bibnamefont {Retin{\`o}}}, \bibinfo {author} {\bibfnamefont {O.}~\bibnamefont {Le~Contel}}, \bibinfo {author} {\bibfnamefont {D.}~\bibnamefont {Perrone}}, \bibinfo {author} {\bibfnamefont {O.~W.}\ \bibnamefont {Roberts}}, \bibinfo {author} {\bibfnamefont {J.~T.}\ \bibnamefont {Coburn}}, \bibinfo {author} {\bibfnamefont {V.}~\bibnamefont {Panebianco}}, \bibinfo {author} {\bibfnamefont {F.}~\bibnamefont {Valentini}}, \bibinfo {author} {\bibfnamefont {S.}~\bibnamefont {Perri}},  \emph {et~al.},\ }\href@noop {} {\bibfield  {journal} {\bibinfo  {journal} {Physical Review Letters}\ }\textbf {\bibinfo {volume} {122}},\ \bibinfo {pages} {035102} (\bibinfo {year} {2019})}\BibitemShut {NoStop}%
\bibitem [{\citenamefont {Wilder}\ \emph {et~al.}(2016)\citenamefont {Wilder}, \citenamefont {Ergun}, \citenamefont {Schwartz}, \citenamefont {Newman}, \citenamefont {Eriksson}, \citenamefont {Stawarz}, \citenamefont {Goldman}, \citenamefont {Goodrich}, \citenamefont {Gershman}, \citenamefont {Malaspina} \emph {et~al.}}]{wilder2016observations}%
  \BibitemOpen
  \bibfield  {author} {\bibinfo {author} {\bibfnamefont {F.}~\bibnamefont {Wilder}}, \bibinfo {author} {\bibfnamefont {R.}~\bibnamefont {Ergun}}, \bibinfo {author} {\bibfnamefont {S.}~\bibnamefont {Schwartz}}, \bibinfo {author} {\bibfnamefont {D.}~\bibnamefont {Newman}}, \bibinfo {author} {\bibfnamefont {S.}~\bibnamefont {Eriksson}}, \bibinfo {author} {\bibfnamefont {J.}~\bibnamefont {Stawarz}}, \bibinfo {author} {\bibfnamefont {M.}~\bibnamefont {Goldman}}, \bibinfo {author} {\bibfnamefont {K.}~\bibnamefont {Goodrich}}, \bibinfo {author} {\bibfnamefont {D.}~\bibnamefont {Gershman}}, \bibinfo {author} {\bibfnamefont {D.}~\bibnamefont {Malaspina}},  \emph {et~al.},\ }\href@noop {} {\bibfield  {journal} {\bibinfo  {journal} {Geophysical Research Letters}\ }\textbf {\bibinfo {volume} {43}},\ \bibinfo {pages} {8859} (\bibinfo {year} {2016})}\BibitemShut {NoStop}%
\bibitem [{1()}]{1}%
  \BibitemOpen
  \href@noop {} {}\bibinfo {note} {See Supplemental Materials for the repetitive crossings of the magnetopause boundary layer over the $2.5$ hours interval, which include Refs.~\cite {Burch16,Russell16:mms,Ergun16:ssr,Lindqvist16,Pollock16:mms}.}\BibitemShut {Stop}%
\bibitem [{2()}]{2}%
  \BibitemOpen
  \href@noop {} {}\bibinfo {note} {See the four-spacecraft interferometry analysis of Alfv\'en waves in Supplemental Materials, which include Ref.~\cite {paschmann2000issi}}\BibitemShut {NoStop}%
\bibitem [{\citenamefont {{Russell}}\ \emph {et~al.}(2016)\citenamefont {{Russell}}, \citenamefont {{Anderson}}, \citenamefont {{Baumjohann}}, \citenamefont {{Bromund}}, \citenamefont {{Dearborn}}, \citenamefont {{Fischer}}, \citenamefont {{Le}}, \citenamefont {{Leinweber}}, \citenamefont {{Leneman}}, \citenamefont {{Magnes}}, \citenamefont {{Means}}, \citenamefont {{Moldwin}}, \citenamefont {{Nakamura}}, \citenamefont {{Pierce}}, \citenamefont {{Plaschke}}, \citenamefont {{Rowe}}, \citenamefont {{Slavin}}, \citenamefont {{Strangeway}}, \citenamefont {{Torbert}}, \citenamefont {{Hagen}}, \citenamefont {{Jernej}}, \citenamefont {{Valavanoglou}},\ and\ \citenamefont {{Richter}}}]{Russell16:mms}%
  \BibitemOpen
  \bibfield  {author} {\bibinfo {author} {\bibfnamefont {C.~T.}\ \bibnamefont {{Russell}}}, \bibinfo {author} {\bibfnamefont {B.~J.}\ \bibnamefont {{Anderson}}}, \bibinfo {author} {\bibfnamefont {W.}~\bibnamefont {{Baumjohann}}}, \bibinfo {author} {\bibfnamefont {K.~R.}\ \bibnamefont {{Bromund}}}, \bibinfo {author} {\bibfnamefont {D.}~\bibnamefont {{Dearborn}}}, \bibinfo {author} {\bibfnamefont {D.}~\bibnamefont {{Fischer}}}, \bibinfo {author} {\bibfnamefont {G.}~\bibnamefont {{Le}}}, \bibinfo {author} {\bibfnamefont {H.~K.}\ \bibnamefont {{Leinweber}}}, \bibinfo {author} {\bibfnamefont {D.}~\bibnamefont {{Leneman}}}, \bibinfo {author} {\bibfnamefont {W.}~\bibnamefont {{Magnes}}}, \bibinfo {author} {\bibfnamefont {J.~D.}\ \bibnamefont {{Means}}}, \bibinfo {author} {\bibfnamefont {M.~B.}\ \bibnamefont {{Moldwin}}}, \bibinfo {author} {\bibfnamefont {R.}~\bibnamefont {{Nakamura}}}, \bibinfo {author} {\bibfnamefont {D.}~\bibnamefont {{Pierce}}}, \bibinfo {author} {\bibfnamefont {F.}~\bibnamefont {{Plaschke}}},
  \bibinfo {author} {\bibfnamefont {K.~M.}\ \bibnamefont {{Rowe}}}, \bibinfo {author} {\bibfnamefont {J.~A.}\ \bibnamefont {{Slavin}}}, \bibinfo {author} {\bibfnamefont {R.~J.}\ \bibnamefont {{Strangeway}}}, \bibinfo {author} {\bibfnamefont {R.}~\bibnamefont {{Torbert}}}, \bibinfo {author} {\bibfnamefont {C.}~\bibnamefont {{Hagen}}}, \bibinfo {author} {\bibfnamefont {I.}~\bibnamefont {{Jernej}}}, \bibinfo {author} {\bibfnamefont {A.}~\bibnamefont {{Valavanoglou}}}, \ and\ \bibinfo {author} {\bibfnamefont {I.}~\bibnamefont {{Richter}}},\ }\href {\doibase 10.1007/s11214-014-0057-3} {\bibfield  {journal} {\bibinfo  {journal} {\ssr}\ }\textbf {\bibinfo {volume} {199}},\ \bibinfo {pages} {189} (\bibinfo {year} {2016})}\BibitemShut {NoStop}%
\bibitem [{\citenamefont {{Pollock}}\ \emph {et~al.}(2016)\citenamefont {{Pollock}}, \citenamefont {{Moore}}, \citenamefont {{Jacques}}, \citenamefont {{Burch}}, \citenamefont {{Gliese}}, \citenamefont {{Saito}}, \citenamefont {{Omoto}}, \citenamefont {{Avanov}}, \citenamefont {{Barrie}}, \citenamefont {{Coffey}}, \citenamefont {{Dorelli}}, \citenamefont {{Gershman}}, \citenamefont {{Giles}}, \citenamefont {{Rosnack}}, \citenamefont {{Salo}}, \citenamefont {{Yokota}}, \citenamefont {{Adrian}}, \citenamefont {{Aoustin}}, \citenamefont {{Auletti}}, \citenamefont {{Aung}}, \citenamefont {{Bigio}}, \citenamefont {{Cao}}, \citenamefont {{Chandler}}, \citenamefont {{Chornay}}, \citenamefont {{Christian}}, \citenamefont {{Clark}}, \citenamefont {{Collinson}}, \citenamefont {{Corris}}, \citenamefont {{De Los Santos}}, \citenamefont {{Devlin}}, \citenamefont {{Diaz}}, \citenamefont {{Dickerson}}, \citenamefont {{Dickson}}, \citenamefont {{Diekmann}}, \citenamefont {{Diggs}}, \citenamefont {{Duncan}}, \citenamefont
  {{Figueroa-Vinas}}, \citenamefont {{Firman}}, \citenamefont {{Freeman}}, \citenamefont {{Galassi}}, \citenamefont {{Garcia}}, \citenamefont {{Goodhart}}, \citenamefont {{Guererro}}, \citenamefont {{Hageman}}, \citenamefont {{Hanley}}, \citenamefont {{Hemminger}}, \citenamefont {{Holland}}, \citenamefont {{Hutchins}}, \citenamefont {{James}}, \citenamefont {{Jones}}, \citenamefont {{Kreisler}}, \citenamefont {{Kujawski}}, \citenamefont {{Lavu}}, \citenamefont {{Lobell}}, \citenamefont {{LeCompte}}, \citenamefont {{Lukemire}}, \citenamefont {{MacDonald}}, \citenamefont {{Mariano}}, \citenamefont {{Mukai}}, \citenamefont {{Narayanan}}, \citenamefont {{Nguyan}}, \citenamefont {{Onizuka}}, \citenamefont {{Paterson}}, \citenamefont {{Persyn}}, \citenamefont {{Piepgrass}}, \citenamefont {{Cheney}}, \citenamefont {{Rager}}, \citenamefont {{Raghuram}}, \citenamefont {{Ramil}}, \citenamefont {{Reichenthal}}, \citenamefont {{Rodriguez}}, \citenamefont {{Rouzaud}}, \citenamefont {{Rucker}}, \citenamefont {{Saito}},
  \citenamefont {{Samara}}, \citenamefont {{Sauvaud}}, \citenamefont {{Schuster}}, \citenamefont {{Shappirio}}, \citenamefont {{Shelton}}, \citenamefont {{Sher}}, \citenamefont {{Smith}}, \citenamefont {{Smith}}, \citenamefont {{Smith}}, \citenamefont {{Steinfeld}}, \citenamefont {{Szymkiewicz}}, \citenamefont {{Tanimoto}}, \citenamefont {{Taylor}}, \citenamefont {{Tucker}}, \citenamefont {{Tull}}, \citenamefont {{Uhl}}, \citenamefont {{Vloet}}, \citenamefont {{Walpole}}, \citenamefont {{Weidner}}, \citenamefont {{White}}, \citenamefont {{Winkert}}, \citenamefont {{Yeh}},\ and\ \citenamefont {{Zeuch}}}]{Pollock16:mms}%
  \BibitemOpen
  \bibfield  {author} {\bibinfo {author} {\bibfnamefont {C.}~\bibnamefont {{Pollock}}}, \bibinfo {author} {\bibfnamefont {T.}~\bibnamefont {{Moore}}}, \bibinfo {author} {\bibfnamefont {A.}~\bibnamefont {{Jacques}}}, \bibinfo {author} {\bibfnamefont {J.}~\bibnamefont {{Burch}}}, \bibinfo {author} {\bibfnamefont {U.}~\bibnamefont {{Gliese}}}, \bibinfo {author} {\bibfnamefont {Y.}~\bibnamefont {{Saito}}}, \bibinfo {author} {\bibfnamefont {T.}~\bibnamefont {{Omoto}}}, \bibinfo {author} {\bibfnamefont {L.}~\bibnamefont {{Avanov}}}, \bibinfo {author} {\bibfnamefont {A.}~\bibnamefont {{Barrie}}}, \bibinfo {author} {\bibfnamefont {V.}~\bibnamefont {{Coffey}}}, \bibinfo {author} {\bibfnamefont {J.}~\bibnamefont {{Dorelli}}}, \bibinfo {author} {\bibfnamefont {D.}~\bibnamefont {{Gershman}}}, \bibinfo {author} {\bibfnamefont {B.}~\bibnamefont {{Giles}}}, \bibinfo {author} {\bibfnamefont {T.}~\bibnamefont {{Rosnack}}}, \bibinfo {author} {\bibfnamefont {C.}~\bibnamefont {{Salo}}}, \bibinfo {author} {\bibfnamefont
  {S.}~\bibnamefont {{Yokota}}}, \bibinfo {author} {\bibfnamefont {M.}~\bibnamefont {{Adrian}}}, \bibinfo {author} {\bibfnamefont {C.}~\bibnamefont {{Aoustin}}}, \bibinfo {author} {\bibfnamefont {C.}~\bibnamefont {{Auletti}}}, \bibinfo {author} {\bibfnamefont {S.}~\bibnamefont {{Aung}}}, \bibinfo {author} {\bibfnamefont {V.}~\bibnamefont {{Bigio}}}, \bibinfo {author} {\bibfnamefont {N.}~\bibnamefont {{Cao}}}, \bibinfo {author} {\bibfnamefont {M.}~\bibnamefont {{Chandler}}}, \bibinfo {author} {\bibfnamefont {D.}~\bibnamefont {{Chornay}}}, \bibinfo {author} {\bibfnamefont {K.}~\bibnamefont {{Christian}}}, \bibinfo {author} {\bibfnamefont {G.}~\bibnamefont {{Clark}}}, \bibinfo {author} {\bibfnamefont {G.}~\bibnamefont {{Collinson}}}, \bibinfo {author} {\bibfnamefont {T.}~\bibnamefont {{Corris}}}, \bibinfo {author} {\bibfnamefont {A.}~\bibnamefont {{De Los Santos}}}, \bibinfo {author} {\bibfnamefont {R.}~\bibnamefont {{Devlin}}}, \bibinfo {author} {\bibfnamefont {T.}~\bibnamefont {{Diaz}}}, \bibinfo {author}
  {\bibfnamefont {T.}~\bibnamefont {{Dickerson}}}, \bibinfo {author} {\bibfnamefont {C.}~\bibnamefont {{Dickson}}}, \bibinfo {author} {\bibfnamefont {A.}~\bibnamefont {{Diekmann}}}, \bibinfo {author} {\bibfnamefont {F.}~\bibnamefont {{Diggs}}}, \bibinfo {author} {\bibfnamefont {C.}~\bibnamefont {{Duncan}}}, \bibinfo {author} {\bibfnamefont {A.}~\bibnamefont {{Figueroa-Vinas}}}, \bibinfo {author} {\bibfnamefont {C.}~\bibnamefont {{Firman}}}, \bibinfo {author} {\bibfnamefont {M.}~\bibnamefont {{Freeman}}}, \bibinfo {author} {\bibfnamefont {N.}~\bibnamefont {{Galassi}}}, \bibinfo {author} {\bibfnamefont {K.}~\bibnamefont {{Garcia}}}, \bibinfo {author} {\bibfnamefont {G.}~\bibnamefont {{Goodhart}}}, \bibinfo {author} {\bibfnamefont {D.}~\bibnamefont {{Guererro}}}, \bibinfo {author} {\bibfnamefont {J.}~\bibnamefont {{Hageman}}}, \bibinfo {author} {\bibfnamefont {J.}~\bibnamefont {{Hanley}}}, \bibinfo {author} {\bibfnamefont {E.}~\bibnamefont {{Hemminger}}}, \bibinfo {author} {\bibfnamefont {M.}~\bibnamefont
  {{Holland}}}, \bibinfo {author} {\bibfnamefont {M.}~\bibnamefont {{Hutchins}}}, \bibinfo {author} {\bibfnamefont {T.}~\bibnamefont {{James}}}, \bibinfo {author} {\bibfnamefont {W.}~\bibnamefont {{Jones}}}, \bibinfo {author} {\bibfnamefont {S.}~\bibnamefont {{Kreisler}}}, \bibinfo {author} {\bibfnamefont {J.}~\bibnamefont {{Kujawski}}}, \bibinfo {author} {\bibfnamefont {V.}~\bibnamefont {{Lavu}}}, \bibinfo {author} {\bibfnamefont {J.}~\bibnamefont {{Lobell}}}, \bibinfo {author} {\bibfnamefont {E.}~\bibnamefont {{LeCompte}}}, \bibinfo {author} {\bibfnamefont {A.}~\bibnamefont {{Lukemire}}}, \bibinfo {author} {\bibfnamefont {E.}~\bibnamefont {{MacDonald}}}, \bibinfo {author} {\bibfnamefont {A.}~\bibnamefont {{Mariano}}}, \bibinfo {author} {\bibfnamefont {T.}~\bibnamefont {{Mukai}}}, \bibinfo {author} {\bibfnamefont {K.}~\bibnamefont {{Narayanan}}}, \bibinfo {author} {\bibfnamefont {Q.}~\bibnamefont {{Nguyan}}}, \bibinfo {author} {\bibfnamefont {M.}~\bibnamefont {{Onizuka}}}, \bibinfo {author} {\bibfnamefont
  {W.}~\bibnamefont {{Paterson}}}, \bibinfo {author} {\bibfnamefont {S.}~\bibnamefont {{Persyn}}}, \bibinfo {author} {\bibfnamefont {B.}~\bibnamefont {{Piepgrass}}}, \bibinfo {author} {\bibfnamefont {F.}~\bibnamefont {{Cheney}}}, \bibinfo {author} {\bibfnamefont {A.}~\bibnamefont {{Rager}}}, \bibinfo {author} {\bibfnamefont {T.}~\bibnamefont {{Raghuram}}}, \bibinfo {author} {\bibfnamefont {A.}~\bibnamefont {{Ramil}}}, \bibinfo {author} {\bibfnamefont {L.}~\bibnamefont {{Reichenthal}}}, \bibinfo {author} {\bibfnamefont {H.}~\bibnamefont {{Rodriguez}}}, \bibinfo {author} {\bibfnamefont {J.}~\bibnamefont {{Rouzaud}}}, \bibinfo {author} {\bibfnamefont {A.}~\bibnamefont {{Rucker}}}, \bibinfo {author} {\bibfnamefont {Y.}~\bibnamefont {{Saito}}}, \bibinfo {author} {\bibfnamefont {M.}~\bibnamefont {{Samara}}}, \bibinfo {author} {\bibfnamefont {J.-A.}\ \bibnamefont {{Sauvaud}}}, \bibinfo {author} {\bibfnamefont {D.}~\bibnamefont {{Schuster}}}, \bibinfo {author} {\bibfnamefont {M.}~\bibnamefont {{Shappirio}}}, \bibinfo
  {author} {\bibfnamefont {K.}~\bibnamefont {{Shelton}}}, \bibinfo {author} {\bibfnamefont {D.}~\bibnamefont {{Sher}}}, \bibinfo {author} {\bibfnamefont {D.}~\bibnamefont {{Smith}}}, \bibinfo {author} {\bibfnamefont {K.}~\bibnamefont {{Smith}}}, \bibinfo {author} {\bibfnamefont {S.}~\bibnamefont {{Smith}}}, \bibinfo {author} {\bibfnamefont {D.}~\bibnamefont {{Steinfeld}}}, \bibinfo {author} {\bibfnamefont {R.}~\bibnamefont {{Szymkiewicz}}}, \bibinfo {author} {\bibfnamefont {K.}~\bibnamefont {{Tanimoto}}}, \bibinfo {author} {\bibfnamefont {J.}~\bibnamefont {{Taylor}}}, \bibinfo {author} {\bibfnamefont {C.}~\bibnamefont {{Tucker}}}, \bibinfo {author} {\bibfnamefont {K.}~\bibnamefont {{Tull}}}, \bibinfo {author} {\bibfnamefont {A.}~\bibnamefont {{Uhl}}}, \bibinfo {author} {\bibfnamefont {J.}~\bibnamefont {{Vloet}}}, \bibinfo {author} {\bibfnamefont {P.}~\bibnamefont {{Walpole}}}, \bibinfo {author} {\bibfnamefont {S.}~\bibnamefont {{Weidner}}}, \bibinfo {author} {\bibfnamefont {D.}~\bibnamefont {{White}}},
  \bibinfo {author} {\bibfnamefont {G.}~\bibnamefont {{Winkert}}}, \bibinfo {author} {\bibfnamefont {P.-S.}\ \bibnamefont {{Yeh}}}, \ and\ \bibinfo {author} {\bibfnamefont {M.}~\bibnamefont {{Zeuch}}},\ }\href {\doibase 10.1007/s11214-016-0245-4} {\bibfield  {journal} {\bibinfo  {journal} {\ssr}\ }\textbf {\bibinfo {volume} {199}},\ \bibinfo {pages} {331} (\bibinfo {year} {2016})}\BibitemShut {NoStop}%
\bibitem [{\citenamefont {{Lindqvist}}\ \emph {et~al.}(2016)\citenamefont {{Lindqvist}}, \citenamefont {{Olsson}}, \citenamefont {{Torbert}}, \citenamefont {{King}}, \citenamefont {{Granoff}}, \citenamefont {{Rau}}, \citenamefont {{Needell}}, \citenamefont {{Turco}}, \citenamefont {{Dors}}, \citenamefont {{Beckman}}, \citenamefont {{Macri}}, \citenamefont {{Frost}}, \citenamefont {{Salwen}}, \citenamefont {{Eriksson}}, \citenamefont {{{\AA}hl{\'e}n}}, \citenamefont {{Khotyaintsev}}, \citenamefont {{Porter}}, \citenamefont {{Lappalainen}}, \citenamefont {{Ergun}}, \citenamefont {{Wermeer}},\ and\ \citenamefont {{Tucker}}}]{Lindqvist16}%
  \BibitemOpen
  \bibfield  {author} {\bibinfo {author} {\bibfnamefont {P.-A.}\ \bibnamefont {{Lindqvist}}}, \bibinfo {author} {\bibfnamefont {G.}~\bibnamefont {{Olsson}}}, \bibinfo {author} {\bibfnamefont {R.~B.}\ \bibnamefont {{Torbert}}}, \bibinfo {author} {\bibfnamefont {B.}~\bibnamefont {{King}}}, \bibinfo {author} {\bibfnamefont {M.}~\bibnamefont {{Granoff}}}, \bibinfo {author} {\bibfnamefont {D.}~\bibnamefont {{Rau}}}, \bibinfo {author} {\bibfnamefont {G.}~\bibnamefont {{Needell}}}, \bibinfo {author} {\bibfnamefont {S.}~\bibnamefont {{Turco}}}, \bibinfo {author} {\bibfnamefont {I.}~\bibnamefont {{Dors}}}, \bibinfo {author} {\bibfnamefont {P.}~\bibnamefont {{Beckman}}}, \bibinfo {author} {\bibfnamefont {J.}~\bibnamefont {{Macri}}}, \bibinfo {author} {\bibfnamefont {C.}~\bibnamefont {{Frost}}}, \bibinfo {author} {\bibfnamefont {J.}~\bibnamefont {{Salwen}}}, \bibinfo {author} {\bibfnamefont {A.}~\bibnamefont {{Eriksson}}}, \bibinfo {author} {\bibfnamefont {L.}~\bibnamefont {{{\AA}hl{\'e}n}}}, \bibinfo {author}
  {\bibfnamefont {Y.~V.}\ \bibnamefont {{Khotyaintsev}}}, \bibinfo {author} {\bibfnamefont {J.}~\bibnamefont {{Porter}}}, \bibinfo {author} {\bibfnamefont {K.}~\bibnamefont {{Lappalainen}}}, \bibinfo {author} {\bibfnamefont {R.~E.}\ \bibnamefont {{Ergun}}}, \bibinfo {author} {\bibfnamefont {W.}~\bibnamefont {{Wermeer}}}, \ and\ \bibinfo {author} {\bibfnamefont {S.}~\bibnamefont {{Tucker}}},\ }\href {\doibase 10.1007/s11214-014-0116-9} {\bibfield  {journal} {\bibinfo  {journal} {\ssr}\ }\textbf {\bibinfo {volume} {199}},\ \bibinfo {pages} {137} (\bibinfo {year} {2016})}\BibitemShut {NoStop}%
\bibitem [{\citenamefont {{Ergun}}\ \emph {et~al.}(2016)\citenamefont {{Ergun}}, \citenamefont {{Tucker}}, \citenamefont {{Westfall}}, \citenamefont {{Goodrich}}, \citenamefont {{Malaspina}}, \citenamefont {{Summers}}, \citenamefont {{Wallace}}, \citenamefont {{Karlsson}}, \citenamefont {{Mack}}, \citenamefont {{Brennan}}, \citenamefont {{Pyke}}, \citenamefont {{Withnell}}, \citenamefont {{Torbert}}, \citenamefont {{Macri}}, \citenamefont {{Rau}}, \citenamefont {{Dors}}, \citenamefont {{Needell}}, \citenamefont {{Lindqvist}}, \citenamefont {{Olsson}},\ and\ \citenamefont {{Cully}}}]{Ergun16:ssr}%
  \BibitemOpen
  \bibfield  {author} {\bibinfo {author} {\bibfnamefont {R.~E.}\ \bibnamefont {{Ergun}}}, \bibinfo {author} {\bibfnamefont {S.}~\bibnamefont {{Tucker}}}, \bibinfo {author} {\bibfnamefont {J.}~\bibnamefont {{Westfall}}}, \bibinfo {author} {\bibfnamefont {K.~A.}\ \bibnamefont {{Goodrich}}}, \bibinfo {author} {\bibfnamefont {D.~M.}\ \bibnamefont {{Malaspina}}}, \bibinfo {author} {\bibfnamefont {D.}~\bibnamefont {{Summers}}}, \bibinfo {author} {\bibfnamefont {J.}~\bibnamefont {{Wallace}}}, \bibinfo {author} {\bibfnamefont {M.}~\bibnamefont {{Karlsson}}}, \bibinfo {author} {\bibfnamefont {J.}~\bibnamefont {{Mack}}}, \bibinfo {author} {\bibfnamefont {N.}~\bibnamefont {{Brennan}}}, \bibinfo {author} {\bibfnamefont {B.}~\bibnamefont {{Pyke}}}, \bibinfo {author} {\bibfnamefont {P.}~\bibnamefont {{Withnell}}}, \bibinfo {author} {\bibfnamefont {R.}~\bibnamefont {{Torbert}}}, \bibinfo {author} {\bibfnamefont {J.}~\bibnamefont {{Macri}}}, \bibinfo {author} {\bibfnamefont {D.}~\bibnamefont {{Rau}}}, \bibinfo {author}
  {\bibfnamefont {I.}~\bibnamefont {{Dors}}}, \bibinfo {author} {\bibfnamefont {J.}~\bibnamefont {{Needell}}}, \bibinfo {author} {\bibfnamefont {P.-A.}\ \bibnamefont {{Lindqvist}}}, \bibinfo {author} {\bibfnamefont {G.}~\bibnamefont {{Olsson}}}, \ and\ \bibinfo {author} {\bibfnamefont {C.~M.}\ \bibnamefont {{Cully}}},\ }\href {\doibase 10.1007/s11214-014-0115-x} {\bibfield  {journal} {\bibinfo  {journal} {\ssr}\ }\textbf {\bibinfo {volume} {199}},\ \bibinfo {pages} {167} (\bibinfo {year} {2016})}\BibitemShut {NoStop}%
\bibitem [{3()}]{3}%
  \BibitemOpen
  \href@noop {} {}\bibinfo {note} {See Supplemental Materials for the detailed inter-antenna interferometry analysis, which include Refs.~\cite {vasko2018solitary,vasko2020nature,Sonnerup68}}\BibitemShut {NoStop}%
\bibitem [{\citenamefont {Bowers}\ \emph {et~al.}(2008)\citenamefont {Bowers}, \citenamefont {Albright}, \citenamefont {Yin}, \citenamefont {Bergen},\ and\ \citenamefont {Kwan}}]{bowers2008ultrahigh}%
  \BibitemOpen
  \bibfield  {author} {\bibinfo {author} {\bibfnamefont {K.~J.}\ \bibnamefont {Bowers}}, \bibinfo {author} {\bibfnamefont {B.~J.}\ \bibnamefont {Albright}}, \bibinfo {author} {\bibfnamefont {L.}~\bibnamefont {Yin}}, \bibinfo {author} {\bibfnamefont {B.}~\bibnamefont {Bergen}}, \ and\ \bibinfo {author} {\bibfnamefont {T.~J.}\ \bibnamefont {Kwan}},\ }\href@noop {} {\bibfield  {journal} {\bibinfo  {journal} {Physics of Plasmas}\ }\textbf {\bibinfo {volume} {15}} (\bibinfo {year} {2008})}\BibitemShut {NoStop}%
\bibitem [{\citenamefont {Le}\ \emph {et~al.}(2023)\citenamefont {Le}, \citenamefont {Stanier}, \citenamefont {Yin}, \citenamefont {Wetherton}, \citenamefont {Keenan},\ and\ \citenamefont {Albright}}]{le2023hybrid}%
  \BibitemOpen
  \bibfield  {author} {\bibinfo {author} {\bibfnamefont {A.}~\bibnamefont {Le}}, \bibinfo {author} {\bibfnamefont {A.}~\bibnamefont {Stanier}}, \bibinfo {author} {\bibfnamefont {L.}~\bibnamefont {Yin}}, \bibinfo {author} {\bibfnamefont {B.}~\bibnamefont {Wetherton}}, \bibinfo {author} {\bibfnamefont {B.}~\bibnamefont {Keenan}}, \ and\ \bibinfo {author} {\bibfnamefont {B.}~\bibnamefont {Albright}},\ }\href@noop {} {\bibfield  {journal} {\bibinfo  {journal} {Physics of Plasmas}\ }\textbf {\bibinfo {volume} {30}} (\bibinfo {year} {2023})}\BibitemShut {NoStop}%
\bibitem [{\citenamefont {Cohen}\ and\ \citenamefont {Kulsrud}(1974)}]{cohen1974nonlinear}%
  \BibitemOpen
  \bibfield  {author} {\bibinfo {author} {\bibfnamefont {R.~H.}\ \bibnamefont {Cohen}}\ and\ \bibinfo {author} {\bibfnamefont {R.~M.}\ \bibnamefont {Kulsrud}},\ }\href@noop {} {\bibfield  {journal} {\bibinfo  {journal} {The Physics of Fluids}\ }\textbf {\bibinfo {volume} {17}},\ \bibinfo {pages} {2215} (\bibinfo {year} {1974})}\BibitemShut {NoStop}%
\bibitem [{\citenamefont {Gonz{\'a}lez}\ \emph {et~al.}(2021)\citenamefont {Gonz{\'a}lez}, \citenamefont {Tenerani}, \citenamefont {Matteini}, \citenamefont {Hellinger},\ and\ \citenamefont {Velli}}]{gonzalez2021proton}%
  \BibitemOpen
  \bibfield  {author} {\bibinfo {author} {\bibfnamefont {C.}~\bibnamefont {Gonz{\'a}lez}}, \bibinfo {author} {\bibfnamefont {A.}~\bibnamefont {Tenerani}}, \bibinfo {author} {\bibfnamefont {L.}~\bibnamefont {Matteini}}, \bibinfo {author} {\bibfnamefont {P.}~\bibnamefont {Hellinger}}, \ and\ \bibinfo {author} {\bibfnamefont {M.}~\bibnamefont {Velli}},\ }\href@noop {} {\bibfield  {journal} {\bibinfo  {journal} {The Astrophysical Journal Letters}\ }\textbf {\bibinfo {volume} {914}},\ \bibinfo {pages} {L36} (\bibinfo {year} {2021})}\BibitemShut {NoStop}%
\bibitem [{\citenamefont {Buneman}(1959)}]{buneman1959dissipation}%
  \BibitemOpen
  \bibfield  {author} {\bibinfo {author} {\bibfnamefont {O.}~\bibnamefont {Buneman}},\ }\href@noop {} {\bibfield  {journal} {\bibinfo  {journal} {Physical Review}\ }\textbf {\bibinfo {volume} {115}},\ \bibinfo {pages} {503} (\bibinfo {year} {1959})}\BibitemShut {NoStop}%
\bibitem [{\citenamefont {Davidson}\ \emph {et~al.}(1970)\citenamefont {Davidson}, \citenamefont {Krall}, \citenamefont {Papadopoulos},\ and\ \citenamefont {Shanny}}]{davidson1970electron}%
  \BibitemOpen
  \bibfield  {author} {\bibinfo {author} {\bibfnamefont {R.}~\bibnamefont {Davidson}}, \bibinfo {author} {\bibfnamefont {N.}~\bibnamefont {Krall}}, \bibinfo {author} {\bibfnamefont {K.}~\bibnamefont {Papadopoulos}}, \ and\ \bibinfo {author} {\bibfnamefont {R.}~\bibnamefont {Shanny}},\ }\href@noop {} {\bibfield  {journal} {\bibinfo  {journal} {Physical Review Letters}\ }\textbf {\bibinfo {volume} {24}},\ \bibinfo {pages} {579} (\bibinfo {year} {1970})}\BibitemShut {NoStop}%
\bibitem [{4()}]{4}%
  \BibitemOpen
  \href@noop {} {}\bibinfo {note} {IAWs are allowed to propagate in the hybrid-kinetic scheme. However, the Debye-scale IAWs cannot be resolved in this scheme, because there is no other intrinsic, physical spatial scale in the system smaller than the ion inertial length.}\BibitemShut {Stop}%
\bibitem [{\citenamefont {Gonz{\'a}lez}\ \emph {et~al.}(2023)\citenamefont {Gonz{\'a}lez}, \citenamefont {Innocenti},\ and\ \citenamefont {Tenerani}}]{gonzalez2023particle}%
  \BibitemOpen
  \bibfield  {author} {\bibinfo {author} {\bibfnamefont {C.}~\bibnamefont {Gonz{\'a}lez}}, \bibinfo {author} {\bibfnamefont {M.~E.}\ \bibnamefont {Innocenti}}, \ and\ \bibinfo {author} {\bibfnamefont {A.}~\bibnamefont {Tenerani}},\ }\href@noop {} {\bibfield  {journal} {\bibinfo  {journal} {Journal of Plasma Physics}\ }\textbf {\bibinfo {volume} {89}},\ \bibinfo {pages} {905890208} (\bibinfo {year} {2023})}\BibitemShut {NoStop}%
\bibitem [{\citenamefont {Holloway}\ and\ \citenamefont {Dorning}(1991)}]{holloway1991undamped}%
  \BibitemOpen
  \bibfield  {author} {\bibinfo {author} {\bibfnamefont {J.~P.}\ \bibnamefont {Holloway}}\ and\ \bibinfo {author} {\bibfnamefont {J.}~\bibnamefont {Dorning}},\ }\href@noop {} {\bibfield  {journal} {\bibinfo  {journal} {Physical Review A}\ }\textbf {\bibinfo {volume} {44}},\ \bibinfo {pages} {3856} (\bibinfo {year} {1991})}\BibitemShut {NoStop}%
\bibitem [{\citenamefont {Valentini}\ \emph {et~al.}(2006)\citenamefont {Valentini}, \citenamefont {O’Neil},\ and\ \citenamefont {Dubin}}]{valentini2006excitation}%
  \BibitemOpen
  \bibfield  {author} {\bibinfo {author} {\bibfnamefont {F.}~\bibnamefont {Valentini}}, \bibinfo {author} {\bibfnamefont {T.~M.}\ \bibnamefont {O’Neil}}, \ and\ \bibinfo {author} {\bibfnamefont {D.~H.}\ \bibnamefont {Dubin}},\ }\href@noop {} {\bibfield  {journal} {\bibinfo  {journal} {Physics of plasmas}\ }\textbf {\bibinfo {volume} {13}} (\bibinfo {year} {2006})}\BibitemShut {NoStop}%
\bibitem [{\citenamefont {{An}}\ \emph {et~al.}(2019)\citenamefont {{An}}, \citenamefont {{Li}}, \citenamefont {{Bortnik}}, \citenamefont {{Decyk}}, \citenamefont {{Kletzing}},\ and\ \citenamefont {{Hospodarsky}}}]{An19}%
  \BibitemOpen
  \bibfield  {author} {\bibinfo {author} {\bibfnamefont {X.}~\bibnamefont {{An}}}, \bibinfo {author} {\bibfnamefont {J.}~\bibnamefont {{Li}}}, \bibinfo {author} {\bibfnamefont {J.}~\bibnamefont {{Bortnik}}}, \bibinfo {author} {\bibfnamefont {V.}~\bibnamefont {{Decyk}}}, \bibinfo {author} {\bibfnamefont {C.}~\bibnamefont {{Kletzing}}}, \ and\ \bibinfo {author} {\bibfnamefont {G.}~\bibnamefont {{Hospodarsky}}},\ }\href {\doibase 10.1103/PhysRevLett.122.045101} {\bibfield  {journal} {\bibinfo  {journal} {\prl}\ }\textbf {\bibinfo {volume} {122}},\ \bibinfo {eid} {045101} (\bibinfo {year} {2019})}\BibitemShut {NoStop}%
\bibitem [{\citenamefont {{An}}\ \emph {et~al.}(2021)\citenamefont {{An}}, \citenamefont {{Bortnik}},\ and\ \citenamefont {{Zhang}}}]{An21:kaw}%
  \BibitemOpen
  \bibfield  {author} {\bibinfo {author} {\bibfnamefont {X.}~\bibnamefont {{An}}}, \bibinfo {author} {\bibfnamefont {J.}~\bibnamefont {{Bortnik}}}, \ and\ \bibinfo {author} {\bibfnamefont {X.-J.}\ \bibnamefont {{Zhang}}},\ }\href {\doibase 10.1029/2020JA028643} {\bibfield  {journal} {\bibinfo  {journal} {Journal of Geophysical Research (Space Physics)}\ }\textbf {\bibinfo {volume} {126}},\ \bibinfo {eid} {e28643} (\bibinfo {year} {2021})}\BibitemShut {NoStop}%
\bibitem [{\citenamefont {{G{\'e}not}}\ \emph {et~al.}(2001)\citenamefont {{G{\'e}not}}, \citenamefont {{Louarn}},\ and\ \citenamefont {{Mottez}}}]{Genot01}%
  \BibitemOpen
  \bibfield  {author} {\bibinfo {author} {\bibfnamefont {V.}~\bibnamefont {{G{\'e}not}}}, \bibinfo {author} {\bibfnamefont {P.}~\bibnamefont {{Louarn}}}, \ and\ \bibinfo {author} {\bibfnamefont {F.}~\bibnamefont {{Mottez}}},\ }\href {\doibase 10.1029/2001JA000076} {\bibfield  {journal} {\bibinfo  {journal} {\jgr}\ }\textbf {\bibinfo {volume} {106}},\ \bibinfo {pages} {29633} (\bibinfo {year} {2001})}\BibitemShut {NoStop}%
\bibitem [{\citenamefont {{G{\'e}not}}\ \emph {et~al.}(2004)\citenamefont {{G{\'e}not}}, \citenamefont {{Louarn}},\ and\ \citenamefont {{Mottez}}}]{Genot04}%
  \BibitemOpen
  \bibfield  {author} {\bibinfo {author} {\bibfnamefont {V.}~\bibnamefont {{G{\'e}not}}}, \bibinfo {author} {\bibfnamefont {P.}~\bibnamefont {{Louarn}}}, \ and\ \bibinfo {author} {\bibfnamefont {F.}~\bibnamefont {{Mottez}}},\ }\href {\doibase 10.5194/angeo-22-2081-2004} {\bibfield  {journal} {\bibinfo  {journal} {Annales Geophysicae}\ }\textbf {\bibinfo {volume} {22}},\ \bibinfo {pages} {2081} (\bibinfo {year} {2004})}\BibitemShut {NoStop}%
\bibitem [{\citenamefont {Mottez}\ \emph {et~al.}(1992)\citenamefont {Mottez}, \citenamefont {Chanteur},\ and\ \citenamefont {Roux}}]{mottez1992filamentation}%
  \BibitemOpen
  \bibfield  {author} {\bibinfo {author} {\bibfnamefont {F.}~\bibnamefont {Mottez}}, \bibinfo {author} {\bibfnamefont {G.}~\bibnamefont {Chanteur}}, \ and\ \bibinfo {author} {\bibfnamefont {A.}~\bibnamefont {Roux}},\ }\href@noop {} {\bibfield  {journal} {\bibinfo  {journal} {Journal of Geophysical Research: Space Physics}\ }\textbf {\bibinfo {volume} {97}},\ \bibinfo {pages} {10801} (\bibinfo {year} {1992})}\BibitemShut {NoStop}%
\bibitem [{\citenamefont {Gedalin}\ \emph {et~al.}(2010)\citenamefont {Gedalin}, \citenamefont {Medvedev}, \citenamefont {Spitkovsky}, \citenamefont {Krasnoselskikh}, \citenamefont {Balikhin}, \citenamefont {Vaivads},\ and\ \citenamefont {Perri}}]{gedalin2010growth}%
  \BibitemOpen
  \bibfield  {author} {\bibinfo {author} {\bibfnamefont {M.}~\bibnamefont {Gedalin}}, \bibinfo {author} {\bibfnamefont {M.}~\bibnamefont {Medvedev}}, \bibinfo {author} {\bibfnamefont {A.}~\bibnamefont {Spitkovsky}}, \bibinfo {author} {\bibfnamefont {V.}~\bibnamefont {Krasnoselskikh}}, \bibinfo {author} {\bibfnamefont {M.}~\bibnamefont {Balikhin}}, \bibinfo {author} {\bibfnamefont {A.}~\bibnamefont {Vaivads}}, \ and\ \bibinfo {author} {\bibfnamefont {S.}~\bibnamefont {Perri}},\ }\href@noop {} {\bibfield  {journal} {\bibinfo  {journal} {Physics of Plasmas}\ }\textbf {\bibinfo {volume} {17}} (\bibinfo {year} {2010})}\BibitemShut {NoStop}%
\bibitem [{\citenamefont {Gurnett}\ \emph {et~al.}(1979)\citenamefont {Gurnett}, \citenamefont {Marsch}, \citenamefont {Pilipp}, \citenamefont {Schwenn},\ and\ \citenamefont {Rosenbauer}}]{gurnett1979ion}%
  \BibitemOpen
  \bibfield  {author} {\bibinfo {author} {\bibfnamefont {D.}~\bibnamefont {Gurnett}}, \bibinfo {author} {\bibfnamefont {E.}~\bibnamefont {Marsch}}, \bibinfo {author} {\bibfnamefont {W.}~\bibnamefont {Pilipp}}, \bibinfo {author} {\bibfnamefont {R.}~\bibnamefont {Schwenn}}, \ and\ \bibinfo {author} {\bibfnamefont {H.}~\bibnamefont {Rosenbauer}},\ }\href@noop {} {\bibfield  {journal} {\bibinfo  {journal} {Journal of Geophysical Research: Space Physics}\ }\textbf {\bibinfo {volume} {84}},\ \bibinfo {pages} {2029} (\bibinfo {year} {1979})}\BibitemShut {NoStop}%
\bibitem [{\citenamefont {Kamaletdinov}\ \emph {et~al.}(2022)\citenamefont {Kamaletdinov}, \citenamefont {Vasko}, \citenamefont {Artemyev}, \citenamefont {Wang},\ and\ \citenamefont {Mozer}}]{kamaletdinov2022quantifying}%
  \BibitemOpen
  \bibfield  {author} {\bibinfo {author} {\bibfnamefont {S.}~\bibnamefont {Kamaletdinov}}, \bibinfo {author} {\bibfnamefont {I.}~\bibnamefont {Vasko}}, \bibinfo {author} {\bibfnamefont {A.}~\bibnamefont {Artemyev}}, \bibinfo {author} {\bibfnamefont {R.}~\bibnamefont {Wang}}, \ and\ \bibinfo {author} {\bibfnamefont {F.}~\bibnamefont {Mozer}},\ }\href@noop {} {\bibfield  {journal} {\bibinfo  {journal} {Physics of Plasmas}\ }\textbf {\bibinfo {volume} {29}} (\bibinfo {year} {2022})}\BibitemShut {NoStop}%
\bibitem [{\citenamefont {Kamaletdinov}\ \emph {et~al.}(2024)\citenamefont {Kamaletdinov}, \citenamefont {Vasko},\ and\ \citenamefont {Artemyev}}]{kamaletdinov2024nonlinear}%
  \BibitemOpen
  \bibfield  {author} {\bibinfo {author} {\bibfnamefont {S.~R.}\ \bibnamefont {Kamaletdinov}}, \bibinfo {author} {\bibfnamefont {I.~Y.}\ \bibnamefont {Vasko}}, \ and\ \bibinfo {author} {\bibfnamefont {A.~V.}\ \bibnamefont {Artemyev}},\ }\href@noop {} {\bibfield  {journal} {\bibinfo  {journal} {Journal of Plasma Physics}\ }\textbf {\bibinfo {volume} {90}},\ \bibinfo {pages} {905900201} (\bibinfo {year} {2024})}\BibitemShut {NoStop}%
\bibitem [{\citenamefont {Coroniti}\ and\ \citenamefont {Eviatar}(1977)}]{coroniti1977magnetic}%
  \BibitemOpen
  \bibfield  {author} {\bibinfo {author} {\bibfnamefont {F.}~\bibnamefont {Coroniti}}\ and\ \bibinfo {author} {\bibfnamefont {A.}~\bibnamefont {Eviatar}},\ }\href@noop {} {\bibfield  {journal} {\bibinfo  {journal} {Astrophysical Journal Supplement Series, vol. 33, Feb. 1977, p. 189-210.}\ }\textbf {\bibinfo {volume} {33}},\ \bibinfo {pages} {189} (\bibinfo {year} {1977})}\BibitemShut {NoStop}%
\bibitem [{\citenamefont {Sagdeev}(1979)}]{sagdeev19791976}%
  \BibitemOpen
  \bibfield  {author} {\bibinfo {author} {\bibfnamefont {R.~Z.}\ \bibnamefont {Sagdeev}},\ }\href@noop {} {\bibfield  {journal} {\bibinfo  {journal} {Reviews of Modern Physics}\ }\textbf {\bibinfo {volume} {51}},\ \bibinfo {pages} {1} (\bibinfo {year} {1979})}\BibitemShut {NoStop}%
\bibitem [{\citenamefont {Smith}\ and\ \citenamefont {Priest}(1972)}]{smith1972current}%
  \BibitemOpen
  \bibfield  {author} {\bibinfo {author} {\bibfnamefont {D.~F.}\ \bibnamefont {Smith}}\ and\ \bibinfo {author} {\bibfnamefont {E.}~\bibnamefont {Priest}},\ }\href@noop {} {\bibfield  {journal} {\bibinfo  {journal} {Astrophysical Journal, vol. 176, p. 487}\ }\textbf {\bibinfo {volume} {176}},\ \bibinfo {pages} {487} (\bibinfo {year} {1972})}\BibitemShut {NoStop}%
\bibitem [{\citenamefont {Coppi}\ and\ \citenamefont {Friedland}(1971)}]{coppi1971processes}%
  \BibitemOpen
  \bibfield  {author} {\bibinfo {author} {\bibfnamefont {B.}~\bibnamefont {Coppi}}\ and\ \bibinfo {author} {\bibfnamefont {A.~B.}\ \bibnamefont {Friedland}},\ }\href@noop {} {\bibfield  {journal} {\bibinfo  {journal} {Astrophysical Journal, vol. 169, p. 379}\ }\textbf {\bibinfo {volume} {169}},\ \bibinfo {pages} {379} (\bibinfo {year} {1971})}\BibitemShut {NoStop}%
\bibitem [{\citenamefont {{Angelopoulos}}\ \emph {et~al.}(2019)\citenamefont {{Angelopoulos}}, \citenamefont {{Cruce}}, \citenamefont {{Drozdov}}, \citenamefont {{Grimes}}, \citenamefont {{Hatzigeorgiu}}, \citenamefont {{King}}, \citenamefont {{Larson}}, \citenamefont {{Lewis}}, \citenamefont {{McTiernan}}, \citenamefont {{Roberts}}, \citenamefont {{Russell}}, \citenamefont {{Hori}}, \citenamefont {{Kasahara}}, \citenamefont {{Kumamoto}}, \citenamefont {{Matsuoka}}, \citenamefont {{Miyashita}}, \citenamefont {{Miyoshi}}, \citenamefont {{Shinohara}}, \citenamefont {{Teramoto}}, \citenamefont {{Faden}}, \citenamefont {{Halford}}, \citenamefont {{McCarthy}}, \citenamefont {{Millan}}, \citenamefont {{Sample}}, \citenamefont {{Smith}}, \citenamefont {{Woodger}}, \citenamefont {{Masson}}, \citenamefont {{Narock}}, \citenamefont {{Asamura}}, \citenamefont {{Chang}}, \citenamefont {{Chiang}}, \citenamefont {{Kazama}}, \citenamefont {{Keika}}, \citenamefont {{Matsuda}}, \citenamefont {{Segawa}}, \citenamefont
  {{Seki}}, \citenamefont {{Shoji}}, \citenamefont {{Tam}}, \citenamefont {{Umemura}}, \citenamefont {{Wang}}, \citenamefont {{Wang}}, \citenamefont {{Redmon}}, \citenamefont {{Rodriguez}}, \citenamefont {{Singer}}, \citenamefont {{Vandegriff}}, \citenamefont {{Abe}}, \citenamefont {{Nose}}, \citenamefont {{Shinbori}}, \citenamefont {{Tanaka}}, \citenamefont {{UeNo}}, \citenamefont {{Andersson}}, \citenamefont {{Dunn}}, \citenamefont {{Fowler}}, \citenamefont {{Halekas}}, \citenamefont {{Hara}}, \citenamefont {{Harada}}, \citenamefont {{Lee}}, \citenamefont {{Lillis}}, \citenamefont {{Mitchell}}, \citenamefont {{Argall}}, \citenamefont {{Bromund}}, \citenamefont {{Burch}}, \citenamefont {{Cohen}}, \citenamefont {{Galloy}}, \citenamefont {{Giles}}, \citenamefont {{Jaynes}}, \citenamefont {{Le Contel}}, \citenamefont {{Oka}}, \citenamefont {{Phan}}, \citenamefont {{Walsh}}, \citenamefont {{Westlake}}, \citenamefont {{Wilder}}, \citenamefont {{Bale}}, \citenamefont {{Livi}}, \citenamefont {{Pulupa}},
  \citenamefont {{Whittlesey}}, \citenamefont {{DeWolfe}}, \citenamefont {{Harter}}, \citenamefont {{Lucas}}, \citenamefont {{Auster}}, \citenamefont {{Bonnell}}, \citenamefont {{Cully}}, \citenamefont {{Donovan}}, \citenamefont {{Ergun}}, \citenamefont {{Frey}}, \citenamefont {{Jackel}}, \citenamefont {{Keiling}}, \citenamefont {{Korth}}, \citenamefont {{McFadden}}, \citenamefont {{Nishimura}}, \citenamefont {{Plaschke}}, \citenamefont {{Robert}}, \citenamefont {{Turner}}, \citenamefont {{Weygand}}, \citenamefont {{Candey}}, \citenamefont {{Johnson}}, \citenamefont {{Kovalick}}, \citenamefont {{Liu}}, \citenamefont {{McGuire}}, \citenamefont {{Breneman}}, \citenamefont {{Kersten}},\ and\ \citenamefont {{Schroeder}}}]{Angelopoulos19}%
  \BibitemOpen
  \bibfield  {author} {\bibinfo {author} {\bibfnamefont {V.}~\bibnamefont {{Angelopoulos}}}, \bibinfo {author} {\bibfnamefont {P.}~\bibnamefont {{Cruce}}}, \bibinfo {author} {\bibfnamefont {A.}~\bibnamefont {{Drozdov}}}, \bibinfo {author} {\bibfnamefont {E.~W.}\ \bibnamefont {{Grimes}}}, \bibinfo {author} {\bibfnamefont {N.}~\bibnamefont {{Hatzigeorgiu}}}, \bibinfo {author} {\bibfnamefont {D.~A.}\ \bibnamefont {{King}}}, \bibinfo {author} {\bibfnamefont {D.}~\bibnamefont {{Larson}}}, \bibinfo {author} {\bibfnamefont {J.~W.}\ \bibnamefont {{Lewis}}}, \bibinfo {author} {\bibfnamefont {J.~M.}\ \bibnamefont {{McTiernan}}}, \bibinfo {author} {\bibfnamefont {D.~A.}\ \bibnamefont {{Roberts}}}, \bibinfo {author} {\bibfnamefont {C.~L.}\ \bibnamefont {{Russell}}}, \bibinfo {author} {\bibfnamefont {T.}~\bibnamefont {{Hori}}}, \bibinfo {author} {\bibfnamefont {Y.}~\bibnamefont {{Kasahara}}}, \bibinfo {author} {\bibfnamefont {A.}~\bibnamefont {{Kumamoto}}}, \bibinfo {author} {\bibfnamefont {A.}~\bibnamefont {{Matsuoka}}},
  \bibinfo {author} {\bibfnamefont {Y.}~\bibnamefont {{Miyashita}}}, \bibinfo {author} {\bibfnamefont {Y.}~\bibnamefont {{Miyoshi}}}, \bibinfo {author} {\bibfnamefont {I.}~\bibnamefont {{Shinohara}}}, \bibinfo {author} {\bibfnamefont {M.}~\bibnamefont {{Teramoto}}}, \bibinfo {author} {\bibfnamefont {J.~B.}\ \bibnamefont {{Faden}}}, \bibinfo {author} {\bibfnamefont {A.~J.}\ \bibnamefont {{Halford}}}, \bibinfo {author} {\bibfnamefont {M.}~\bibnamefont {{McCarthy}}}, \bibinfo {author} {\bibfnamefont {R.~M.}\ \bibnamefont {{Millan}}}, \bibinfo {author} {\bibfnamefont {J.~G.}\ \bibnamefont {{Sample}}}, \bibinfo {author} {\bibfnamefont {D.~M.}\ \bibnamefont {{Smith}}}, \bibinfo {author} {\bibfnamefont {L.~A.}\ \bibnamefont {{Woodger}}}, \bibinfo {author} {\bibfnamefont {A.}~\bibnamefont {{Masson}}}, \bibinfo {author} {\bibfnamefont {A.~A.}\ \bibnamefont {{Narock}}}, \bibinfo {author} {\bibfnamefont {K.}~\bibnamefont {{Asamura}}}, \bibinfo {author} {\bibfnamefont {T.~F.}\ \bibnamefont {{Chang}}}, \bibinfo {author}
  {\bibfnamefont {C.-Y.}\ \bibnamefont {{Chiang}}}, \bibinfo {author} {\bibfnamefont {Y.}~\bibnamefont {{Kazama}}}, \bibinfo {author} {\bibfnamefont {K.}~\bibnamefont {{Keika}}}, \bibinfo {author} {\bibfnamefont {S.}~\bibnamefont {{Matsuda}}}, \bibinfo {author} {\bibfnamefont {T.}~\bibnamefont {{Segawa}}}, \bibinfo {author} {\bibfnamefont {K.}~\bibnamefont {{Seki}}}, \bibinfo {author} {\bibfnamefont {M.}~\bibnamefont {{Shoji}}}, \bibinfo {author} {\bibfnamefont {S.~W.~Y.}\ \bibnamefont {{Tam}}}, \bibinfo {author} {\bibfnamefont {N.}~\bibnamefont {{Umemura}}}, \bibinfo {author} {\bibfnamefont {B.-J.}\ \bibnamefont {{Wang}}}, \bibinfo {author} {\bibfnamefont {S.-Y.}\ \bibnamefont {{Wang}}}, \bibinfo {author} {\bibfnamefont {R.}~\bibnamefont {{Redmon}}}, \bibinfo {author} {\bibfnamefont {J.~V.}\ \bibnamefont {{Rodriguez}}}, \bibinfo {author} {\bibfnamefont {H.~J.}\ \bibnamefont {{Singer}}}, \bibinfo {author} {\bibfnamefont {J.}~\bibnamefont {{Vandegriff}}}, \bibinfo {author} {\bibfnamefont {S.}~\bibnamefont
  {{Abe}}}, \bibinfo {author} {\bibfnamefont {M.}~\bibnamefont {{Nose}}}, \bibinfo {author} {\bibfnamefont {A.}~\bibnamefont {{Shinbori}}}, \bibinfo {author} {\bibfnamefont {Y.-M.}\ \bibnamefont {{Tanaka}}}, \bibinfo {author} {\bibfnamefont {S.}~\bibnamefont {{UeNo}}}, \bibinfo {author} {\bibfnamefont {L.}~\bibnamefont {{Andersson}}}, \bibinfo {author} {\bibfnamefont {P.}~\bibnamefont {{Dunn}}}, \bibinfo {author} {\bibfnamefont {C.}~\bibnamefont {{Fowler}}}, \bibinfo {author} {\bibfnamefont {J.~S.}\ \bibnamefont {{Halekas}}}, \bibinfo {author} {\bibfnamefont {T.}~\bibnamefont {{Hara}}}, \bibinfo {author} {\bibfnamefont {Y.}~\bibnamefont {{Harada}}}, \bibinfo {author} {\bibfnamefont {C.~O.}\ \bibnamefont {{Lee}}}, \bibinfo {author} {\bibfnamefont {R.}~\bibnamefont {{Lillis}}}, \bibinfo {author} {\bibfnamefont {D.~L.}\ \bibnamefont {{Mitchell}}}, \bibinfo {author} {\bibfnamefont {M.~R.}\ \bibnamefont {{Argall}}}, \bibinfo {author} {\bibfnamefont {K.}~\bibnamefont {{Bromund}}}, \bibinfo {author} {\bibfnamefont
  {J.~L.}\ \bibnamefont {{Burch}}}, \bibinfo {author} {\bibfnamefont {I.~J.}\ \bibnamefont {{Cohen}}}, \bibinfo {author} {\bibfnamefont {M.}~\bibnamefont {{Galloy}}}, \bibinfo {author} {\bibfnamefont {B.}~\bibnamefont {{Giles}}}, \bibinfo {author} {\bibfnamefont {A.~N.}\ \bibnamefont {{Jaynes}}}, \bibinfo {author} {\bibfnamefont {O.}~\bibnamefont {{Le Contel}}}, \bibinfo {author} {\bibfnamefont {M.}~\bibnamefont {{Oka}}}, \bibinfo {author} {\bibfnamefont {T.~D.}\ \bibnamefont {{Phan}}}, \bibinfo {author} {\bibfnamefont {B.~M.}\ \bibnamefont {{Walsh}}}, \bibinfo {author} {\bibfnamefont {J.}~\bibnamefont {{Westlake}}}, \bibinfo {author} {\bibfnamefont {F.~D.}\ \bibnamefont {{Wilder}}}, \bibinfo {author} {\bibfnamefont {S.~D.}\ \bibnamefont {{Bale}}}, \bibinfo {author} {\bibfnamefont {R.}~\bibnamefont {{Livi}}}, \bibinfo {author} {\bibfnamefont {M.}~\bibnamefont {{Pulupa}}}, \bibinfo {author} {\bibfnamefont {P.}~\bibnamefont {{Whittlesey}}}, \bibinfo {author} {\bibfnamefont {A.}~\bibnamefont {{DeWolfe}}},
  \bibinfo {author} {\bibfnamefont {B.}~\bibnamefont {{Harter}}}, \bibinfo {author} {\bibfnamefont {E.}~\bibnamefont {{Lucas}}}, \bibinfo {author} {\bibfnamefont {U.}~\bibnamefont {{Auster}}}, \bibinfo {author} {\bibfnamefont {J.~W.}\ \bibnamefont {{Bonnell}}}, \bibinfo {author} {\bibfnamefont {C.~M.}\ \bibnamefont {{Cully}}}, \bibinfo {author} {\bibfnamefont {E.}~\bibnamefont {{Donovan}}}, \bibinfo {author} {\bibfnamefont {R.~E.}\ \bibnamefont {{Ergun}}}, \bibinfo {author} {\bibfnamefont {H.~U.}\ \bibnamefont {{Frey}}}, \bibinfo {author} {\bibfnamefont {B.}~\bibnamefont {{Jackel}}}, \bibinfo {author} {\bibfnamefont {A.}~\bibnamefont {{Keiling}}}, \bibinfo {author} {\bibfnamefont {H.}~\bibnamefont {{Korth}}}, \bibinfo {author} {\bibfnamefont {J.~P.}\ \bibnamefont {{McFadden}}}, \bibinfo {author} {\bibfnamefont {Y.}~\bibnamefont {{Nishimura}}}, \bibinfo {author} {\bibfnamefont {F.}~\bibnamefont {{Plaschke}}}, \bibinfo {author} {\bibfnamefont {P.}~\bibnamefont {{Robert}}}, \bibinfo {author} {\bibfnamefont
  {D.~L.}\ \bibnamefont {{Turner}}}, \bibinfo {author} {\bibfnamefont {J.~M.}\ \bibnamefont {{Weygand}}}, \bibinfo {author} {\bibfnamefont {R.~M.}\ \bibnamefont {{Candey}}}, \bibinfo {author} {\bibfnamefont {R.~C.}\ \bibnamefont {{Johnson}}}, \bibinfo {author} {\bibfnamefont {T.}~\bibnamefont {{Kovalick}}}, \bibinfo {author} {\bibfnamefont {M.~H.}\ \bibnamefont {{Liu}}}, \bibinfo {author} {\bibfnamefont {R.~E.}\ \bibnamefont {{McGuire}}}, \bibinfo {author} {\bibfnamefont {A.}~\bibnamefont {{Breneman}}}, \bibinfo {author} {\bibfnamefont {K.}~\bibnamefont {{Kersten}}}, \ and\ \bibinfo {author} {\bibfnamefont {P.}~\bibnamefont {{Schroeder}}},\ }\href {\doibase 10.1007/s11214-018-0576-4} {\bibfield  {journal} {\bibinfo  {journal} {\ssr}\ }\textbf {\bibinfo {volume} {215}},\ \bibinfo {eid} {9} (\bibinfo {year} {2019})}\BibitemShut {NoStop}%
\bibitem [{\citenamefont {{Computational and Information Systems Laboratory}}(2024)}]{derecho}%
  \BibitemOpen
  \bibfield  {author} {\bibinfo {author} {\bibnamefont {{Computational and Information Systems Laboratory}}},\ }\href {https://doi.org/10.5065/qx9a-pg09} {\enquote {\bibinfo {title} {Derecho: {HPE} {C}ray {EX} {S}ystem},}\ }\bibinfo {howpublished} {Boulder, CO: National Center for Atmospheric Research} (\bibinfo {year} {2024})\BibitemShut {NoStop}%
\bibitem [{\citenamefont {Paschmann}\ and\ \citenamefont {Schwartz}(2000)}]{paschmann2000issi}%
  \BibitemOpen
  \bibfield  {author} {\bibinfo {author} {\bibfnamefont {G.}~\bibnamefont {Paschmann}}\ and\ \bibinfo {author} {\bibfnamefont {S.}~\bibnamefont {Schwartz}},\ }in\ \href@noop {} {\emph {\bibinfo {booktitle} {Cluster-II workshop multiscale/multipoint plasma measurements}}},\ Vol.\ \bibinfo {volume} {449}\ (\bibinfo {year} {2000})\ p.~\bibinfo {pages} {99}\BibitemShut {NoStop}%
\bibitem [{\citenamefont {Vasko}\ \emph {et~al.}(2018)\citenamefont {Vasko}, \citenamefont {Mozer}, \citenamefont {Krasnoselskikh}, \citenamefont {Artemyev}, \citenamefont {Agapitov}, \citenamefont {Bale}, \citenamefont {Avanov}, \citenamefont {Ergun}, \citenamefont {Giles}, \citenamefont {Lindqvist} \emph {et~al.}}]{vasko2018solitary}%
  \BibitemOpen
  \bibfield  {author} {\bibinfo {author} {\bibfnamefont {I.}~\bibnamefont {Vasko}}, \bibinfo {author} {\bibfnamefont {F.}~\bibnamefont {Mozer}}, \bibinfo {author} {\bibfnamefont {V.}~\bibnamefont {Krasnoselskikh}}, \bibinfo {author} {\bibfnamefont {A.}~\bibnamefont {Artemyev}}, \bibinfo {author} {\bibfnamefont {O.}~\bibnamefont {Agapitov}}, \bibinfo {author} {\bibfnamefont {S.}~\bibnamefont {Bale}}, \bibinfo {author} {\bibfnamefont {L.}~\bibnamefont {Avanov}}, \bibinfo {author} {\bibfnamefont {R.}~\bibnamefont {Ergun}}, \bibinfo {author} {\bibfnamefont {B.}~\bibnamefont {Giles}}, \bibinfo {author} {\bibfnamefont {P.-A.}\ \bibnamefont {Lindqvist}},  \emph {et~al.},\ }\href@noop {} {\bibfield  {journal} {\bibinfo  {journal} {Geophysical Research Letters}\ }\textbf {\bibinfo {volume} {45}},\ \bibinfo {pages} {5809} (\bibinfo {year} {2018})}\BibitemShut {NoStop}%
\bibitem [{\citenamefont {Vasko}\ \emph {et~al.}(2020)\citenamefont {Vasko}, \citenamefont {Wang}, \citenamefont {Mozer}, \citenamefont {Bale},\ and\ \citenamefont {Artemyev}}]{vasko2020nature}%
  \BibitemOpen
  \bibfield  {author} {\bibinfo {author} {\bibfnamefont {I.~Y.}\ \bibnamefont {Vasko}}, \bibinfo {author} {\bibfnamefont {R.}~\bibnamefont {Wang}}, \bibinfo {author} {\bibfnamefont {F.~S.}\ \bibnamefont {Mozer}}, \bibinfo {author} {\bibfnamefont {S.~D.}\ \bibnamefont {Bale}}, \ and\ \bibinfo {author} {\bibfnamefont {A.~V.}\ \bibnamefont {Artemyev}},\ }\href@noop {} {\bibfield  {journal} {\bibinfo  {journal} {Frontiers in Physics}\ }\textbf {\bibinfo {volume} {8}},\ \bibinfo {pages} {156} (\bibinfo {year} {2020})}\BibitemShut {NoStop}%
\bibitem [{\citenamefont {{Sonnerup}}\ and\ \citenamefont {{Cahill}}(1968)}]{Sonnerup68}%
  \BibitemOpen
  \bibfield  {author} {\bibinfo {author} {\bibfnamefont {B.~U.~{\"O}.}\ \bibnamefont {{Sonnerup}}}\ and\ \bibinfo {author} {\bibfnamefont {L.~J.}\ \bibnamefont {{Cahill}}, \bibfnamefont {Jr.}},\ }\href {\doibase 10.1029/JA073i005p01757} {\bibfield  {journal} {\bibinfo  {journal} {\jgr}\ }\textbf {\bibinfo {volume} {73}},\ \bibinfo {pages} {1757} (\bibinfo {year} {1968})}\BibitemShut {NoStop}%
\end{thebibliography}
%

\end{document}


\preprint{}

\title{Supplemental materials for ``Cross-scale energy transfer from fluid-scale Alfv\'en waves to kinetic-scale ion acoustic waves in the Earth's magnetopause boundary layer''}

\author{Xin An}
\email[]{phyax@ucla.edu}
\affiliation{Department of Earth, Planetary, and Space Sciences, University of California, Los Angeles, CA, 90095, USA}

\author{Anton Artemyev}
\affiliation{Department of Earth, Planetary, and Space Sciences, University of California, Los Angeles, CA, 90095, USA}

\author{Vassilis Angelopoulos}
\affiliation{Department of Earth, Planetary, and Space Sciences, University of California, Los Angeles, CA, 90095, USA}

\author{Terry Z. Liu}
\affiliation{Department of Earth, Planetary, and Space Sciences, University of California, Los Angeles, CA, 90095, USA}

\author{Ivan Vasko}
\affiliation{Department of Physics, University of Texas at Dallas, Richardson, TX, 75080, USA}

\author{David Malaspina}
\affiliation{Astrophysical and Planetary Sciences Department, University of Colorado, Boulder, CO, 80305, USA}
\affiliation{Laboratory for Atmospheric and Space Physics, University of Colorado, Boulder, CO, 80303, USA}

\date{\today}

\begin{abstract}
These Supplemental Materials complement the main manuscript by providing additional supporting information. The first section offers the instrumentation details alongside the event overview spanning $2.5$ hours. In the second section, we delve into the characteristics of Alfv\'en waves, leveraging the inter-spacecraft interferometry to assess their propagation and computing spatial gradients of magnetic-field pressure associated with these waves. Finally, the third section investigates the propagation characteristics of ion acoustic waves through inter-antenna interferometry.
\end{abstract}

\maketitle


\section{Instrumentation and event overview}
This study utilizes three instruments aboard the Magnetospheric Multiscale (MMS) constellation \cite{Burch16}: the Fluxgate Magnetometer (FGM), Electric Double Probes (EDP), and Fast Plasma Investigation (FPI). The FGM records three orthogonal components of magnetic field at a sampling rate of $128$\,Samples/second \cite{Russell16:mms}. The EDP, comprising Axial Double Probes (ADP) \cite{Ergun16:ssr} and Spin-Plane Double Probes (SDP) \cite{Lindqvist16}, measures the full vector of electric field both along and perpendicular to the spin axis at a sampling rate of $8192$\,Samples/second. The FPI captures the $3$D velocity-space distribution of electrons from $10$\,eV to $30$\,keV and ions from $10$\,eV to $30$\,keV with respective time resolutions of $30$\,ms, and $150$\,ms \cite{Pollock16:mms}.

\begin{figure}[hptb]
    \centering
    \includegraphics[width=\textwidth]{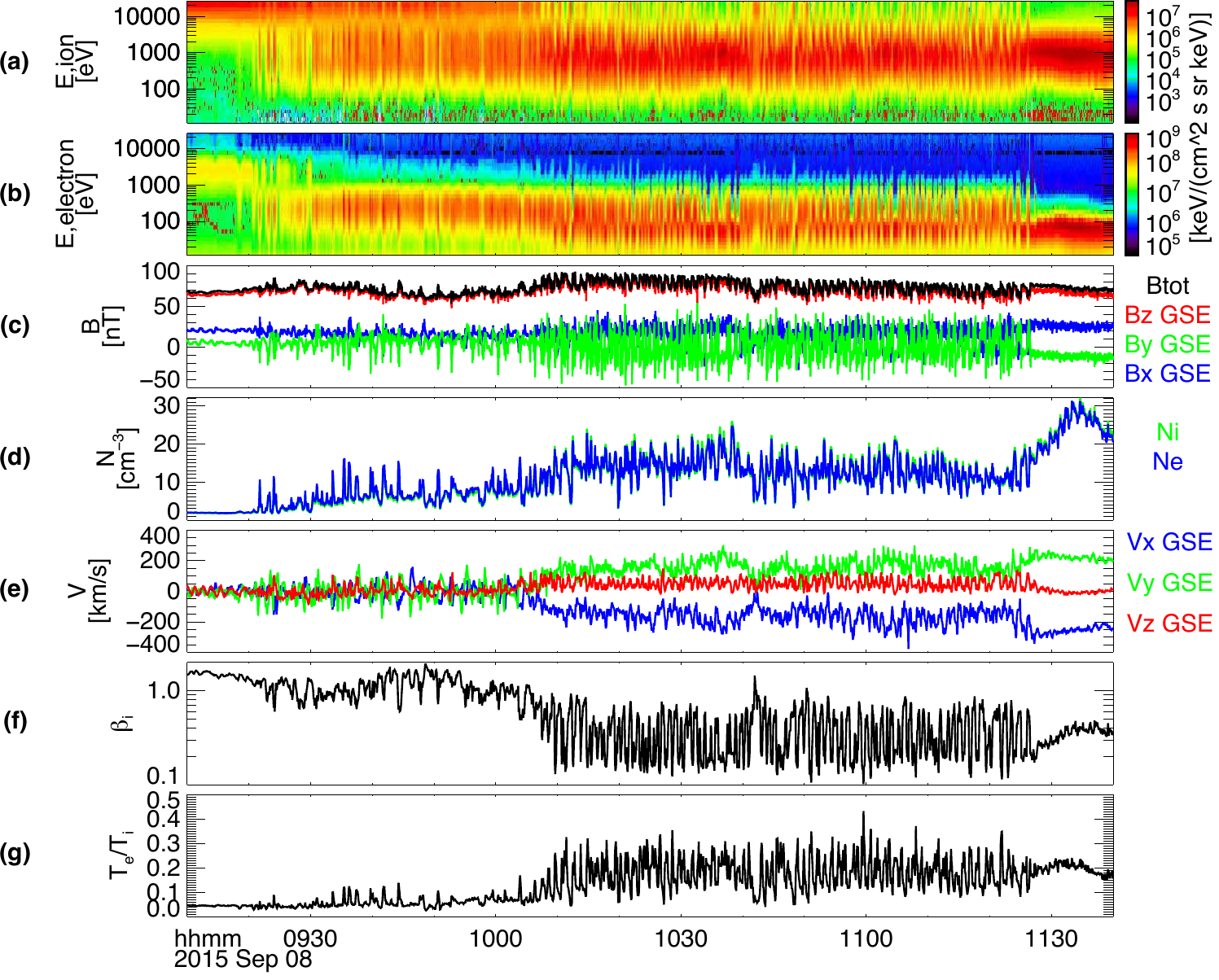}
    \caption{Event overview of the magnetopause boundary layer crossing by MMS1 on 8 September, 2015. The MMS constellation is located around $(X,Y,Z) = (5\,R_\mathrm{E}, 8.8\,R_\mathrm{E}, 0.1\,R_\mathrm{E})$ in the geocentric solar ecliptic (GSE) coordinate, where $R_\mathrm{E}$ is the Earth radius. (a) The ion energy spectrum color-coded by the ion energy flux. (b) The electron energy spectrum color-coded by the electron energy flux. (c) The three components of magnetic field in the GSE coordinate and the total magnetic field magnitude (black). (d) The ion (green) and electron (blue) densities. (e) The three components of ion flow velocity in the GSE coordinate. (f) The ion beta $\beta_i$ defined as the ratio of ion thermal pressure to magnetic field pressure. (g) The electron-to-ion temperature ratio $T_e / T_i$.}
    \label{fig:alfven-overview-2hrs}
\end{figure}

\begin{figure}[hptb]
    \centering
    \includegraphics[width=\textwidth]{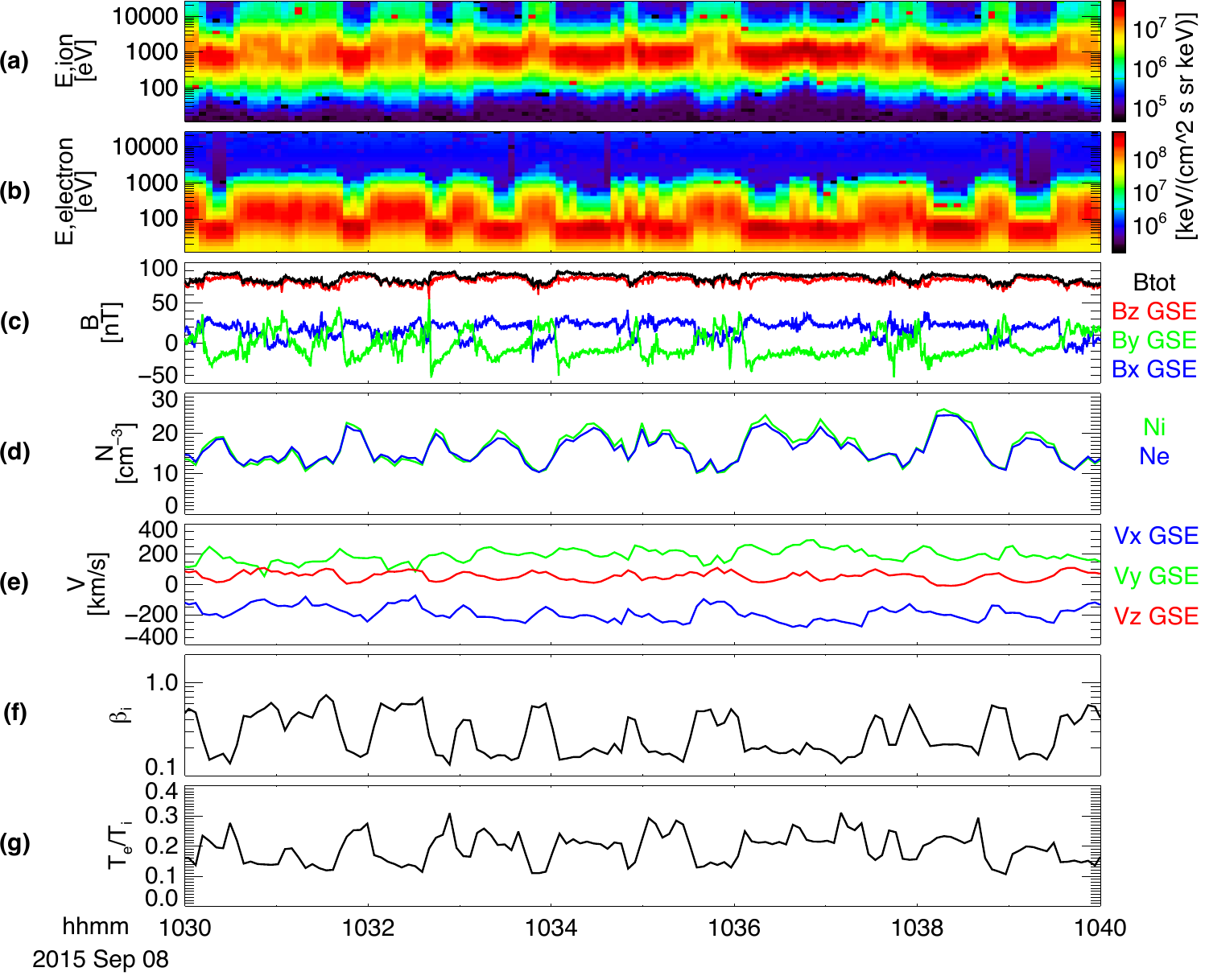}
    \caption{Zoom-in view of the event in Figure \ref{fig:alfven-overview-2hrs} between 10:30--10:40\,UT on 8 September, 2015. The interval shown in the main manuscript is between 10:35:25--10:36:15\,UT.}
    \label{fig:alfven-overview}
\end{figure}

We choose the magnetopause boundary layer as a natural laboratory to study the Alfv\'en-acoustic channeling for ion energy, where plenty of large-amplitude Alfv\'en waves are generated due to solar wind interactions with the magnetosphere. In addition, the ion velocity distributions are better resolved in the boundary layer than in the solar wind. Figure \ref{fig:alfven-overview-2hrs} illustrates a $2.5$-hour traversal of the magnetopause boundary layer by MMS on $8$ September 2015. During this period, surface waves generated by the Kelvin-Helmholtz instability manifest as oscillations in magnetic fields, flow velocities, and ion and electron energy fluxes. A $10$ minute interval of the same event, from 10:30 to 10:40\,UT, is depicted in Figure \ref{fig:alfven-overview}. Notably, the spacecraft alternates between traversing the relatively cold, dense magnetosheath plasma and the hot, tenuous boundary layer, separated by current sheets.

The two pivotal parameters for this study are the ion beta $\beta_i$ and the electron-to-ion temperature ratio $T_e / T_i$. It has been verified that the increase in $\beta_i$ and the decrease in $T_e / T_i$ within the boundary layer are caused by the presence of ion beams compared to the magnetosheath. Excluding beam ions, these parameters are evaluated to be $\beta_i = 0.15$ and $T_e / T_i = 0.25$. This gives the relative locations of Alfv\'en and ion acoustic velocities to the thermal velocity of core ions as $v_{\mathrm{A}} / v_{\mathrm{Ti}} = \sqrt{\beta_i / 2} = 3.7$ and $c_s / v_{\mathrm{Ti}} = \sqrt{(5/3) + (T_e / T_i)} = 1.4$.

\section{Inter-spacecraft interferometry analysis of Alfv\'en waves}
\subsection{Propagation characteristics}
\begin{figure}[hptb]
    \centering
    \includegraphics[width=\textwidth]{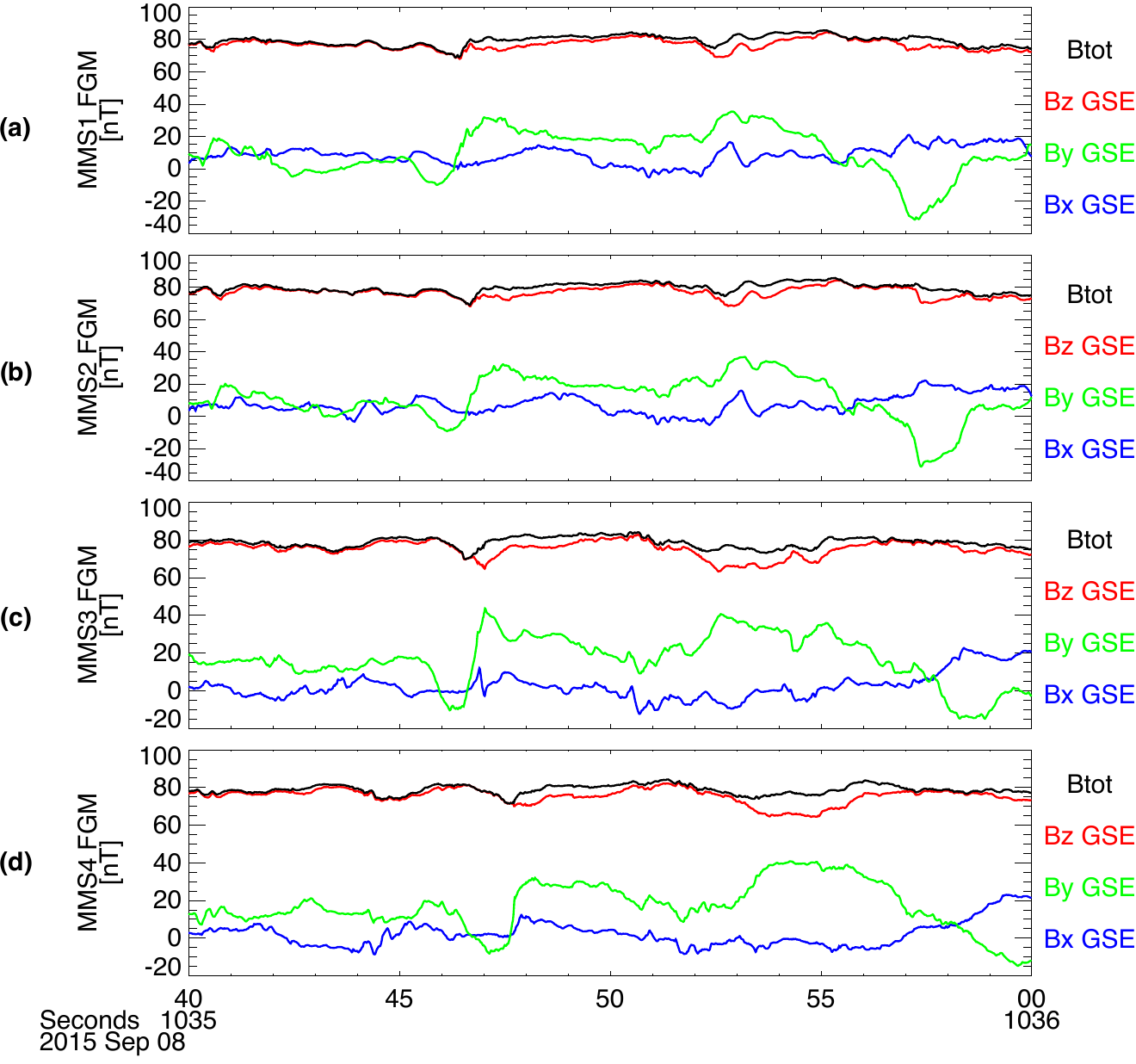}
    \caption{The measurements of magnetic field vector by FGM onboard (a) MMS1, (b) MMS2, (c) MMS3, and (d) MMS4.}
    \label{fig:alfven-b-compare}
\end{figure}

We analyze the propagation characteristics of Alfv\'en waves utilizing the four-satellite interferometry technique. Examination of the magnetic field variations of Alfv\'en waves observed on MMS1--MMS4 reveals distinct time lags, as depicted in Figure \ref{fig:alfven-b-compare}. The propagation velocity of Alfv\'en waves can be obtained by minimizing the sum of the squares of the residual time between each pair of spacecraft, where the residual time refers to the difference between the measured and predicted propagation time.

\begin{figure}[hptb]
    \centering
    \includegraphics[width=0.6\textwidth]{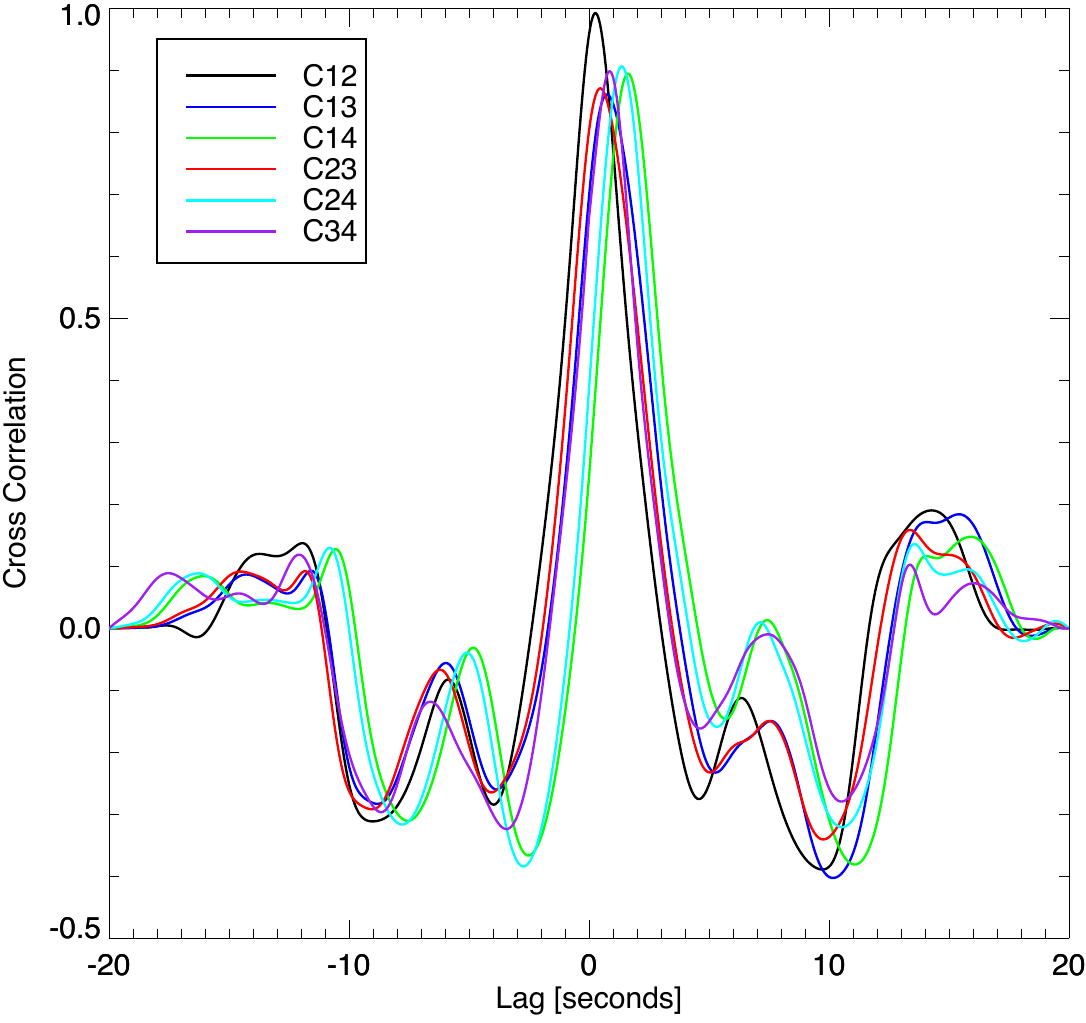}
    \caption{Cross-correlation coefficients of $B_y$ from six pairs of mutually different MMS spacecraft in Figure \ref{fig:alfven-b-compare}. The horizontal axis is the time lag. The time lag corresponding to the peak of cross-correlation coefficient is denoted as the propagation time between each pair of spacecraft.}
    \label{fig:alfven-b-lag}
\end{figure}

The propagation time of the magnetic field $B_y$ between each pair of spacecraft is determined by identifying the peak of their cross-correlation coefficient as a function of time lag, as illustrated in Figure \ref{fig:alfven-b-lag}. $B_y$ is selected for the analysis due to its significant variations during the event. The timings across the four spacecraft are synchronized, and the measured propagation time $t_{\alpha \beta}$ and the corresponding distance $\vert \mathbf{r}_\alpha - \mathbf{r}_\beta \vert$ between each pair of spacecraft $(\alpha, \beta)$ are detailed in Table \ref{tab:alfven-timing}. It is worth noting that the measured propagation time may not strictly satisfy $t_{\alpha \gamma} - t_{\beta \gamma} = t_{\alpha \beta}$. Thus, we determine the propagation velocity by minimizing the cost function defined as:
\begin{equation}\label{eq:cost-timing}
    \mathcal{C} = \sum_{\alpha \neq \beta} \left\vert \frac{\hat{\mathbf{n}} \cdot (\mathbf{r}_\alpha - \mathbf{r}_\beta)}{V} - t_{\alpha \beta} \right\vert^2 ,
\end{equation}
where $\hat{\mathbf{n}}$ represents the unit vector along the propagation direction, $V$ denotes the propagation velocity, and $\sum_{\alpha \neq \beta}$ encompasses all mutually different pairs of spacecraft listed in Table \ref{tab:alfven-timing}. This constitutes a linear optimization problem since the residual time is linear with respect to the vector $\mathbf{m} = \hat{\mathbf{n}} / V$. The solution to the linear optimization problem is given by (see Chapter $12$ in Ref.~\cite{paschmann2000issi}):
\begin{equation}\label{eq:optimization-solution}
    m_l = \frac{1}{4^2} \sum_{\alpha \neq \beta} t_{\alpha \beta} \left( r_{\alpha k} - r_{\beta k} \right) R_{k l}^{-1} ,
\end{equation}
where $r_{\alpha k}$ represents the $k$-th component of the spacecraft position $\mathbf{r}_\alpha$, and the volumetric tensor $R_{kl}$ is defined as:
\begin{equation}\label{eq:volume-tensor}
    R_{k l} = \frac{1}{4} r_{\alpha k} r_{\alpha l} .
\end{equation}
The Einstein summation convention is followed in Equations \eqref{eq:optimization-solution} and \eqref{eq:volume-tensor}. Note that the origin of the coordinate system is chosen at the mean position of the four spacecraft, as the property $\sum_{\alpha} \mathbf{r}_\alpha = 0$ has been used in deriving Equation \eqref{eq:optimization-solution}. From Equation \eqref{eq:optimization-solution}, the magnitude and direction of the propagation velocity in the spacecraft frame in the GSE coordinate are determined to be
\begin{equation}\label{eq:propagation-v-sc}
    V = 103\,\mathrm{km/s}, \hspace{15pt} \hat{\mathbf{n}} = (-0.98, -0.15, 0.1) .
\end{equation}
The dominant wave frequency in the spacecraft frame is $f = 0.05$\,Hz, corresponding to the wavelength $\lambda = V / f = 2059\,\mathrm{km} = 30\,d_i$.

\begin{table}[hptb]
    \centering
    \begin{tabular}{c|c|c}
      \hline
      Pair $(\alpha, \beta)$   &  $t_{\alpha \beta}$\,[seconds]  & $\vert \mathbf{r}_\alpha - \mathbf{r}_\beta \vert$\,[km] \\
      \hline \hline
      MMS2 - MMS1   &  $0.27$  &  $153$  \\
      \hline
      MMS3 - MMS1   &  $0.76$  &  $176$  \\
      \hline
      MMS4 - MMS1   &  $1.63$  &  $183$  \\
      \hline
      MMS3 - MMS2   &  $0.48$  &  $180$  \\
      \hline
      MMS4 - MMS2   &  $1.37$  &  $166$  \\
      \hline
      MMS4 - MMS3   &  $0.87$  &  $200$  \\
      \hline
    \end{tabular}
    \caption{The propagation time $t_{\alpha \beta}$ and distance between six pairs of mutually different spacecraft.}
    \label{tab:alfven-timing}
\end{table}

The propagation velocity in the plasma rest frame can be deduced from that in the spacecraft frame by considering the Doppler effect:
\begin{equation}\label{eq:doppler-effect}
    \omega = \omega' + \mathbf{k} \cdot \mathbf{u} ,
\end{equation}
where $\omega$ is the wave frequency in the spacecraft frame, $\omega'$ is the wave frequency in the plasma rest frame, $\mathbf{k}$ is the corresponding wavenumber, and $\mathbf{u}$ is the mean plasma flow velocity during the event. Dividing both sides of Equation \eqref{eq:doppler-effect} by $k$, we obtain the propagation velocity in the plasma rest frame
\begin{equation}\label{eq:propagation-transform}
    V' = V - \hat{\mathbf{n}} \cdot \mathbf{u} ,
\end{equation}
where $V' = \omega' / k$, $V = \omega / k$, and $\hat{\mathbf{n}} = \mathbf{k} / k$. By substituting the values from Equation \eqref{eq:propagation-v-sc} and $\mathbf{u} = (-114, 151, 85)$\,km/s into Equation \eqref{eq:propagation-transform}, we find the magnitude of the propagation velocity in the plasma frame to be $V' = 5$\,km/s, with the propagation direction remaining $\hat{\mathbf{n}}$. The parallel propagation velocity in the plasma rest frame is $V'_{\parallel} = \omega'/(k \hat{\mathbf{n}} \cdot \hat{\mathbf{b}}) = 539$\,km/s, which is approximately equal to the Alfv\'en velocity.

\subsection{Spatial gradient of magnetic-field pressure}
In Figure 1(c) of the main manuscript, the spatial gradient of the magnetic-field pressure is computed using the four MMS spacecraft. Let $\mathbf{k}$ represent this gradient:
\begin{equation}
    \mathbf{k} = \frac{\partial B^2}{\partial \mathbf{r}} ,
\end{equation}
where $\mathbf{r}$ denotes position. The magnetic pressure $B^2_\alpha$ is simultaneously measured by the four MMS spacecraft ($\alpha$=MMS1, MMS2, MMS3, MMS4). We optimize the spatial gradient of $B^2_\alpha$ by minimizing the cost function:
\begin{equation}\label{eq:cost-grad-b2}
    \mathcal{C} = \sum_{\alpha \neq \beta} \left\vert \mathbf{k} \cdot (\mathbf{r}_\alpha - \mathbf{r}_\beta) - (B^2_\alpha - B^2_\beta) \right\vert^2 .
\end{equation}
Because Equation \eqref{eq:cost-grad-b2} has the same form as Equation \eqref{eq:cost-timing}, with $t_{\alpha \beta}$ replaced by $(B^2_\alpha - B^2_\beta)$, the solution is similar \cite{paschmann2000issi}
\begin{equation}
    k_l = \frac{1}{4^2} \sum_{\alpha \neq \beta} (B^2_\alpha - B^2_\beta) \left( r_{\alpha k} - r_{\beta k} \right) R_{k l}^{-1} ,
\end{equation}
where the volumetric tensor has been defined in Equation \eqref{eq:volume-tensor}.

\section{Inter-antenna interferometry analysis of ion acoustic waves}
\begin{figure}[hptb]
    \centering
    \includegraphics[width=0.75\textwidth]{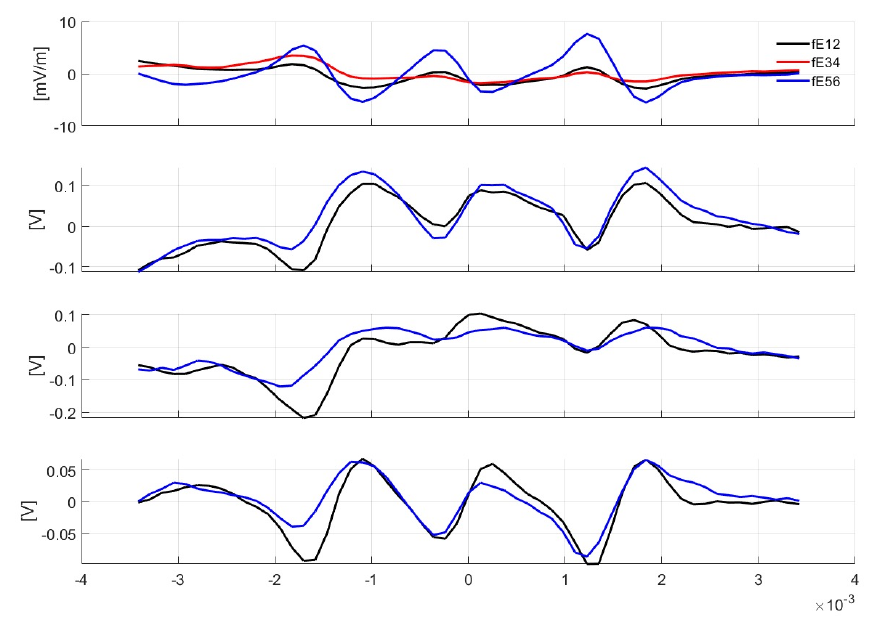}
    \caption{The electric field and associated voltage signals in three pairs of opposing voltage-sensitive probes. (a) The three orthogonal components of electric field in the spin plane and along the spin axis. (b), (c) The voltage signals $V_1$ vs.~$-V_2$ and $V_3$ vs.~$-V_4$ in the two perpendicular directions in the spin plane. (d) The voltage signals $V_5$ vs.~$-V_6$ along the spin axis. The time zero is at 10:35:46.6\,UT on 8, September, 2015. Each division is $1$\,ms on the horizontal axis.}
    \label{fig:iaw-interferometry}
\end{figure}

\begin{figure}[hptb]
    \centering
    \includegraphics[width=0.75\textwidth]{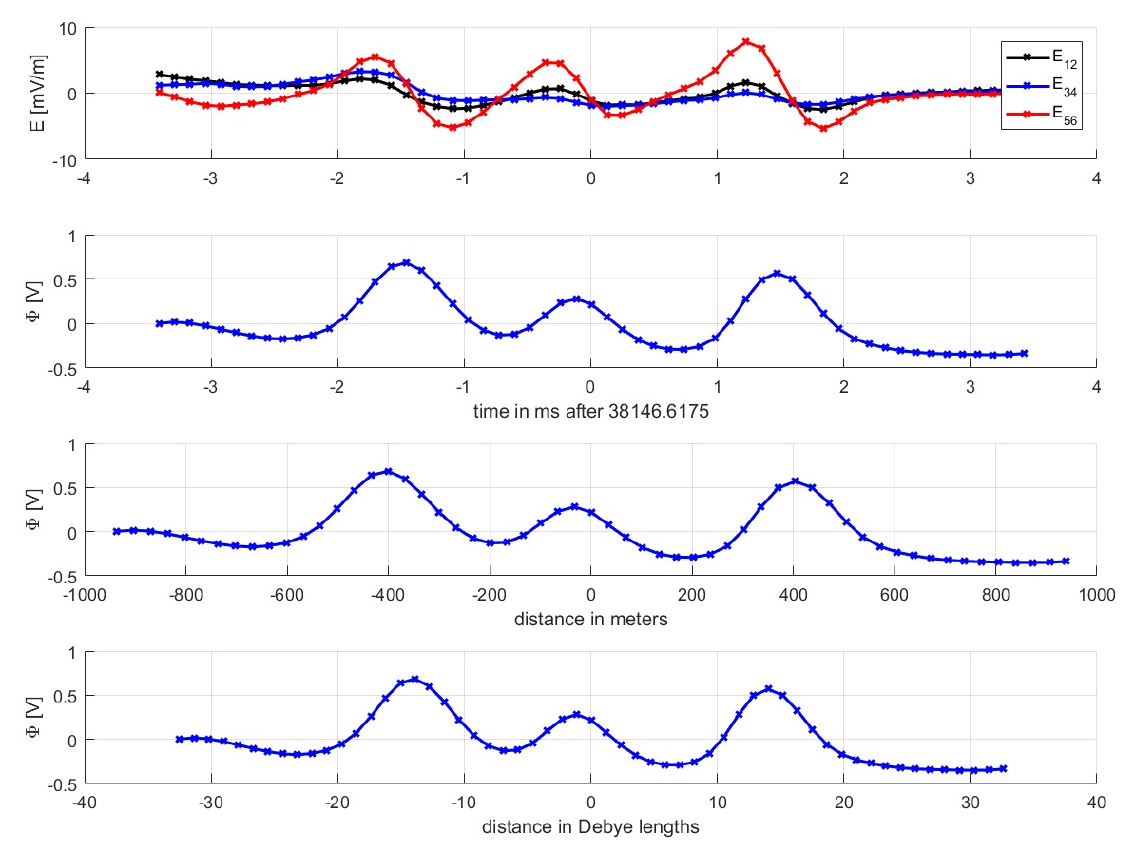}
    \caption{The translation of electric field as a function of time in Figure \ref{fig:iaw-interferometry} to electrostatic potential as a function of position. (a) The temporal profiles of electric field in the spin plane and along the spin axis. (b) The temporal profile of electrostatic potential. (c), (d) The spatial profiles of electrostatic potential in meters and normalized to the local Debye length.}
    \label{fig:iaw-interferometry2}
\end{figure}

We determine the propagation characteristics of ion acoustic waves using the inter-antenna interferometry technique. Figure \ref{fig:iaw-interferometry} shows an example of the three components of electric field in the boom directions, computed as $E_{ij} = c_{ij} (V_i - V_j) / (2 l_{ij})$. Here, $V_i$ and $V_j$ are the voltage signals from two opposing voltage-sensitive probes with a length $l_{ij}$, and the correction factors $c_{ij}$ account for the sensor frequency response and finite boom length \cite{vasko2018solitary}. The voltage signals of $V_1$ vs.~$-V_2$, $V_3$ vs.~$-V_4$, and $V_5$ vs.~$-V_6$ are in the three orthogonal directions. The physical length between the voltage sensors in the spin plane is $2l_{12} = 2l_{34} = 120$\,m, whereas the physical length along the spin axis is $2l_{56} = 29.2$\,m. The time delays between voltage signals between opposing probes allow us to determine the magnitude and direction of the propagation velocity as follows \cite{vasko2018solitary,vasko2020nature}:
\begin{eqnarray}
    \frac{1}{V_s^{2}} = \frac{\Delta t_{12}^2}{l_{12}^2} + \frac{\Delta t_{34}^2}{l_{34}^2} + \frac{\Delta t_{56}^2}{l_{56}^2} & , \\
    \hat{\mathbf{k}} = \left( -\frac{V_s \Delta t_{12}}{l_{12}}, -\frac{V_s \Delta t_{34}}{l_{34}}, -\frac{V_s \Delta t_{56}}{l_{56}} \right) & .
\end{eqnarray}
We find that the propagation velocity is $270$\,km/s in the spacecraft frame and $82$\,km/s in the plasma rest frame, which is approximately equal to the local ion acoustic velocity $90$\,km/s. The propagation angle is within $10^\circ$ relative to the local magnetic field direction and is aligned with the electric field direction, indicating that the observed electric field is approximately a one-dimensional electrostatic structure. The estimated propagation velocity and the one-dimensional nature of the electric field structures allow us to translate their temporal profiles to spatial profiles, as shown in Figure \ref{fig:iaw-interferometry2}. The electrostatic potential is calculated as $\Phi = \int E_l V_s \mathrm{d}t$, where $E_l$ is the dominant electric field determined using minimum variance analysis \cite{Sonnerup68}. The amplitude of the electrostatic potential is about $0.5$\,Volts. The wavelength is $400\,\mathrm{m} \approx 13\,\lambda_D$, where $\lambda_D = 30$\,m is the local Debye length. The propagation characteristics of ion acoustic waves at several other times are listed in Table \ref{tab:iaw-characteristics}. At times, ion acoustic waves also propagate antiparallel to $B_0$, excited by ion beams with $v_\parallel < 0$, likely caused by parts of resonant islands extending to the antiparallel direction.


\begin{table}[hptb]
    \centering
    \begin{tabular}{c|c|c|c|c|c|c|c|c}
    \hline
       & Time [UT] & $V_{\mathrm{sc}}$ [km/s] & $V'$ [km/s] & $V'/c_s$ & $\theta_{kB}$ [$^\circ$] & $\lambda$ [m] & $\lambda/\lambda_D$ & $\Phi_{\max}$ [Volts] \\
     \hline \hline
     MMS1 & 10:35:46.6 & 270 & 82 & 0.4 & 9.5 & 400 & 13 & 0.5 \\
     \hline
     MMS1 & 10:35:48.5 & 62 & -46 & -0.24 & 172 & 200 & 8 & 0.2 \\
     \hline
     MMS1 & 10:35:48.58 & 100 & 14 & 0.07 & 6 & 100 & 4 & 0.2  \\
     \hline
     MMS1 & 10:35:56.3 & -200 & -135 & -0.7 & 174 & 800 & 27 & 0.3 \\
     \hline
     MMS1 & 10:35:54.75 & 124 & 250 & 1.3 & 10 & 100 & 3 & 0.05  \\
     \hline
     MMS2 & 10:35:41.26 & -120 & -50 & -0.26 & 177 & 200 & 10 & 0.4 \\
     \hline
     MMS2 & 10:35:43.3 & -44 & -12 & -0.06 & 175 & 200 & 10 & 0.2 \\
     \hline
     MMS2 & 10:35:54.46 & -53 & -230 & -1.2 & 168 & 50 & 1 & 0.2 \\
     \hline 
     MMS2 & 10:35:55.37 & -79 & -11 & -0.06 & 174 & 200 & 10 & 0.2 \\
     \hline
     MMS2 & 10:35:55.9 & -84 & 23 & 0.12 & 11 & 400 & 30 & 0.2 \\
     \hline
    \end{tabular}
    \caption{Propagation characteristics of ion acoustic waves measured by MMS1 and MMS2. $V_{\mathrm{SC}}$ and $V'$ represent the propagation velocity of ion acoustic waves in the spacecraft and plasma rest frames, respectively. $\theta_{kB}$ represents the angle between the propagation direction and the magnetic field direction. $\lambda$ represents the wavelength. $\Phi_{\max}$ represents the maximum amplitude of the electrostatic potential. $c_s$ is the ion acoustic velocity. $\lambda_D$ is the local Debye length. The propagation characteristics of ion acoustic waves measured by MMS3 and MMS4 are similar to those observed by MMS1 and MM2.}
    \label{tab:iaw-characteristics}
\end{table}




%